\documentclass[a4paper,fleqn]{cas-dc}

\usepackage{subfigure}
\usepackage[authoryear]{natbib}
\usepackage{times}
\usepackage{soul}
\usepackage{subfigure}
\usepackage{url}
\usepackage{multirow}
\usepackage{amssymb}
\usepackage[utf8]{inputenc}
\usepackage[small]{caption}
\usepackage{graphicx}
\usepackage{amsmath}
\usepackage{amsthm}
\usepackage{booktabs}
\usepackage{algorithm}
\usepackage[switch]{lineno}
\usepackage{algorithm}
\usepackage{multirow}
\usepackage{graphicx}
\usepackage{algpseudocode}
\usepackage{microtype}
\usepackage{verbatim}

\def\tsc#1{\csdef{#1}{\textsc{\lowercase{#1}}\xspace}}
\tsc{WGM}
\tsc{QE}
\tsc{EP}
\tsc{PMS}
\tsc{BEC}
\tsc{DE}
\begin{document}

\let\WriteBookmarks\relax
\def\floatpagepagefraction{1}
\def\textpagefraction{.001}
\shorttitle{}

\shortauthors{}

\title [mode = title]{Ternary Spike-based Neuromorphic Signal Processing System}                      
\tnotetext[1]{This work was supported by the National Science Foundation of China under Grant 62106038, and in part by the Sichuan Science and Technology Program under Grant 2023YFG0259.}

\author[1]{Shuai Wang}[style=chinese]
\credit{Conceptualization, Methodology, Experimentation, Writing - Original draft preparation}
\fnmark[1]
\author[1]{Dehao Zhang}[style=chinese]
\credit{Methodology, Investigation}
\author[2]{Ammar~Belatreche}[style=English]
\credit{Methodology, Investigation, Experimentation}
\author[1]{Yichen Xiao}[style=chinese]
\credit{Methodology, Investigation}
\author[1]{Hongyu Qing}[style=chinese]
\credit{Methodology, Investigation, Experimentation}
\author[1]{Wenjie Wei}[style=English]
\credit{Methodology, Investigation, Experimentation}

\author[1]{Malu Zhang}[style=chinese, orcid=0000-0002-2345-0974]
\ead{maluzhang@uestc.edu.cn}
\cormark[1]
\credit{Conceptualization, Supervision, Funding acquisition.}

\author[1]{Yang Yang}[style=chinese]
\credit{Supervision, Investigation}

\address[1]{Department of Computer Science and Engineering, University of Electronic Science and Technology of China, Chengdu, 611731, China}
\address[2]{Department of Computer and Information Sciences, Faculty of Engineering and Environment, Northumbria University, Newcastle upon Tyne NE1 8ST, U.K.}


\begin{abstract}
Deep Neural Networks (DNNs) have been successfully implemented across various signal processing fields, resulting in significant enhancements in performance. However, DNNs generally require substantial computational resources, leading to significant economic costs and posing challenges for their deployment on resource-constrained edge devices.
In this study, we take advantage of spiking neural networks (SNNs) and quantization technologies to develop an energy-efficient and lightweight neuromorphic signal processing system. Our system is characterized by two principal innovations: a threshold-adaptive encoding (TAE) method and a quantized ternary SNN (QT-SNN). The TAE method can efficiently encode time-varying analog signals into sparse ternary spike trains, thereby reducing energy and memory demands for signal processing. QT-SNN, compatible with ternary spike trains from the TAE method, quantifies both membrane potentials and synaptic weights to reduce memory requirements while maintaining performance. Extensive experiments are conducted on two typical signal-processing tasks: speech and electroencephalogram recognition. The results demonstrate that our neuromorphic signal processing system achieves state-of-the-art (SOTA) performance with a 94\% reduced memory requirement. Furthermore, through theoretical energy consumption analysis, our system shows $7.5\times$ energy saving compared to other SNN works. 
The efficiency and efficacy of the proposed system highlight its potential as a promising avenue for energy-efficient signal processing.

\end{abstract}


\maketitle

\section{Introduction}
\label{introduction}


Deep Neural Networks (DNNs) have revolutionized traditional signal processing techniques, achieving higher accuracy in many tasks such as speech recognition~\citep{radford2023robust,liu2023towards,kim2023branchformer} and electroencephalogram (EEG) analysis~\citep{cai2019svm,su2022stanet}. However, with the development of the Internet of Things and edge computing, the substantial computational and memory requirements of DNNs pose challenges for their direct deployment on edge devices. It limits real-time signal processing and immediate decision-making~\citep{troussard2023real, giordano2021accessing, parekh2022review, faisal2019understanding, valsalan2020iot}. Thus, devising more lightweight, energy-efficient, and high-performance intelligent signal processing models remains a challenging problem awaiting resolution.


Spiking Neural Networks (SNNs)~\citep{snn_ref4,snn_ref2,snn_ref3} are inspired by the information transmission mechanisms observed in biological neurons and are considered the third generation of Artificial Neural Networks (ANNs). Spiking neurons compute only upon the arrival of input spikes and remain silent otherwise~\citep{snn_ref1}. Such event-driven mechanism ensures sparse operations and mitigates extensive floating-point multiplication (MAC) operations in SNNs~\citep{caviglia2014asynchronous,event_driven}. 
Thus, when implemented in hardware, SNNs utilizing accumulate (AC) operations exhibit a substantially lower power consumption~\citep{orchard2015converting} compared to MAC-dependent ANNs. In the past years, the signal processing field has seen many innovative and efficient SNN-based solutions~\citep{cai2021eeg,chu2022neuromorphic,safa2021improving}, all achieving satisfactory performance. However, they still face two major bottlenecks.

The first bottleneck lies in the lack of efficient neural encoding methods for signals. 
Some researchers~\citep{wu2018spiking, pan2020efficient, xiao2017spiking} preprocess signals into spectra using Fast Fourier Transform (FFT) and Mel-Frequency Cepstral Coefficients (MFCC), then encode these spectra into spike trains for SNN-based models. However, these preprocessing technologies need massive computing resources, opposing our goal of creating energy-efficient systems. Alternatively, many researchers have adopted the direct encoding method to avoid using preprocessing techniques. This approach directly feeds the raw signal into SNN-based intelligent models, with the model's initial layer as the encoding layer~\citep{weidel2021wavesense,yang2022deep}. However, this approach fails to account for signal characteristics and significantly increases MAC operations in the initial model layer. Therefore, devising more efficient neural signal encoding schemes remains a challenge.

In addition, the complexity and memory requirements of SNNs present another significant bottleneck. 
Recently, an increasing number of high-performance SNNs have been proposed, achieving satisfactory results across many tasks~\citep{zhou2024spikformer,wang2023complex,yao2024spike,zhou2022spikformer}. 
However, these models typically exhibit considerable complexity, necessitating extensive computational resources and memory requirements, rendering them unsuitable for deploying on resource-limited edge devices.
To further exploit the energy efficiency and hardware-friendly advantages of SNNs, many researchers~\citep{hu2021quantized,li2022quantization,sulaiman2020weight,castagnetti2023trainable} have explored quantizing synaptic weights to lower bit-width successfully. However, few works~\citep{yin2023mint} focus on the unique membrane potentials within SNNs, leaving substantial room for improvement. 

In this research, we present a novel ternary spike-based neuromorphic signal processing system to address the challenges previously mentioned. It primarily comprises two innovative components.
First, we propose a threshold-adaptive encoding (TAE) method for effectively encoding raw signals into ternary spike trains and avoiding using high-energy-consuming signal preprocessing techniques.
Second, we propose a dual-scale quantization ternary spiking neural network (QT-SNN). It can further process ternary spike signals from our TAE method, using lower bit-width synapse weights and membrane potential. 
Finally, we validate the performance of our system on two classical signal-processing tasks: keyword recognition and EEG identification. Satisfyingly, compared to other similar efforts, our work achieves state-of-the-art (SOTA) performance with reduced memory usage and lower energy consumption. The principal contributions are summarized as follows:
\begin{itemize}
    \item We propose an innovative threshold-adaptive encoding (TAE) method. This scheme adaptively adjusts thresholds in response to the time-varying characteristics of raw analog signals. This method permits more sparse and efficient signal encoding into ternary spikes, substantially lowering signal transmission's energy and memory requirements.
    \item We propose a dual-scaling quantization ternary spiking neural network (QT-SNN). QT-SNN can directly process ternary spike trains from our TAE method, and further quantize synaptic weights and membrane potential in SNNs to lower bit-width, effectively reducing the memory requirements of SNNs with enhanced performance.
    \item We integrate the TAE method with QT-SNN to develop a ternary spike-based neuromorphic signal processing system. Compared with similar works, our system achieves SOTA performance in both speech and EEG recognition tasks while reducing 94\% of memory footprints and $7.5\times$ energy consumption.
\end{itemize}

\begin{figure*}[htpb] 
\centering
\includegraphics[scale=0.5]{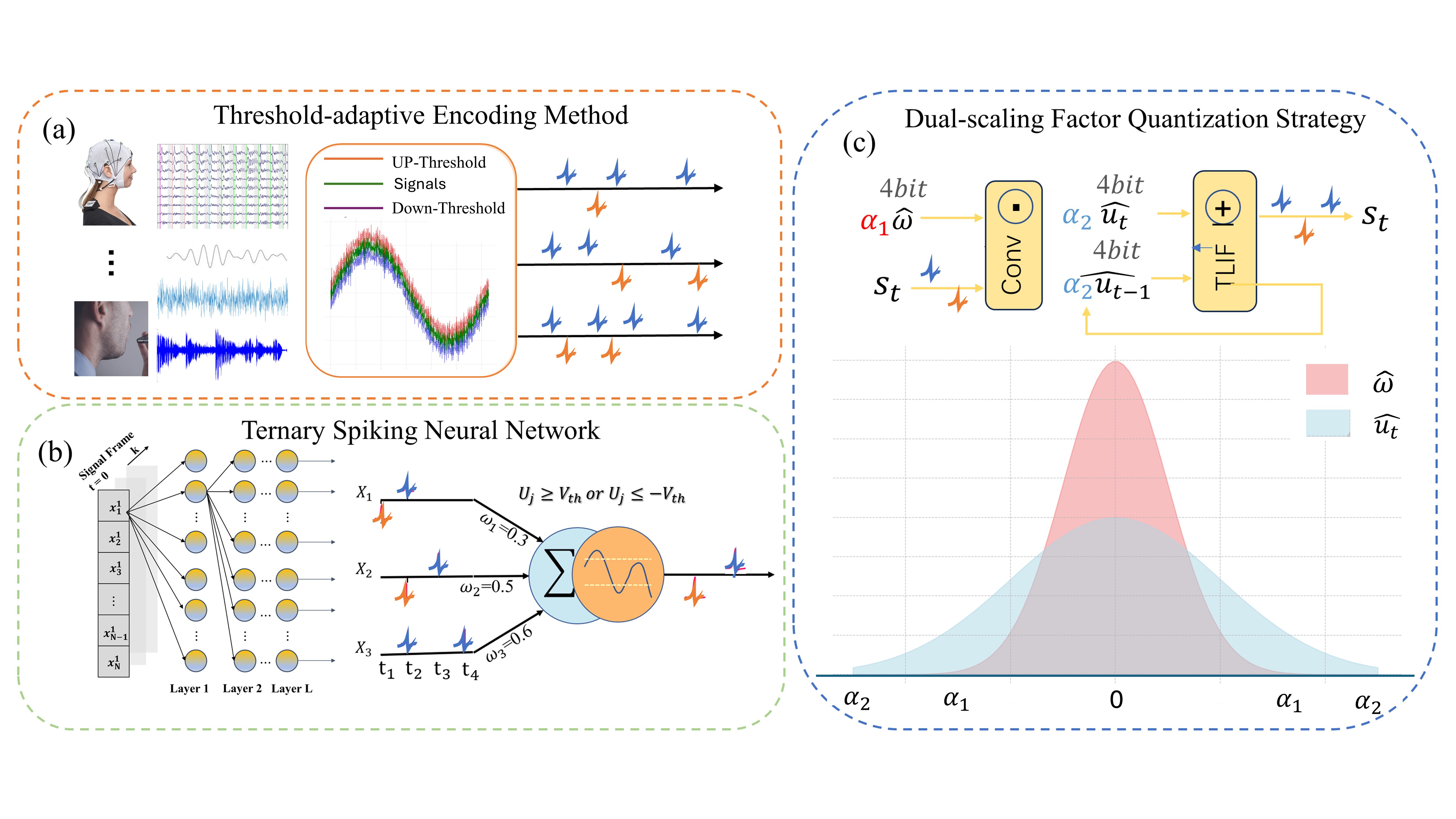}
\caption{Ternary spike-based neuromorphic signal processing system. (a) Threshold-adaptive encoding method. It can efficiently encode analog signal data into ternary spike trains, thereby reducing memory footprints and energy consumption on signal transmission. (b,c) Dual-scaling quantization ternary spiking neural network. It enhances the SNNs' performance and quantizes both synaptic weights and membrane potential to lower bit-width, significantly reducing the network's memory and computational resource requirements.}
\label{Fig555}
\end{figure*}

\section{Related Work}
In this section, we introduce the latest research findings on neural encoding for signals and spiking neural networks. Simultaneously, we analyze the problems and challenges inherent in existing methods.  
\subsection{Neural Encoding for Signals}
Efficient neural coding methods for signals significantly reduce the bandwidth and memory requirements for information transmission. Furthermore, this process provides robust data support for subsequent intelligent SNN-based models. Currently, neural encoding for signals can primarily be divided into two types.

\textbf{Spectrum-based encoding:} Given signals' rapid time-variance and high temporal complexity (e.g., speech signals always have 16,000Hz to 48,000Hz sampling rates), directly analyzing and encoding raw analog signals is highly challenging. 
Consequently, many studies employ MFCC and FFT preprocessing techniques to initially convert analog signals into spectra, and then encode these spectra into spike trains. 
For example, ~\cite{xiao2017spiking} encodes the audio spectrum envelope into spike emission timings through temporal encoding; ~\cite{pan2020efficient} considered the masking characteristics of human auditory perception and applied time-frequency masking to process the spectrum, achieving more sparse neural signal encoding method. However, these methods face two significant issues. First, the current signal preprocessing techniques are resource-intensive and lack energy efficiency. Second, these preprocessing methods incur high latency, hindering their application in fast-decision edge devices.

\textbf{Raw signal-based encoding:} 
To avoid high-resource signal preprocessing techniques, many researchers~\citep{tan2021improved,snn_ref4,snn_ref2} are turning to the DC method. It feeds raw analog signals directly into the SNN-based models, utilizing the model's initial layer for encoding. For example, ~\cite{weidel2021wavesense} employed dilated temporal convolution to process raw speech datasets, achieving accuracy in keyword recognition tasks nearly comparable to ANNs. Subsequently, ~\cite{yang2022deep} enhanced accuracy in the same tasks by employing residual convolutional blocks to process raw speech directly. Nonetheless, the DC method leads to extensive MAC operations in the model's primary layer, compromising the energy efficiency of SNNs.
Recently, some threshold-based encoding methods for signals have been introduced~\citep{kasabov2013dynamic, kasabov2016evolving, dupeyroux2022toolbox} to enable real-time, efficient encoding of raw signals into ternary spike sequences, effectively addressing the limitations of the previous methods. 
However, these methods rely on predetermined thresholds and suffer from significant loss of information during the encoding process. Thus, devising methods for more effective and energy-efficient encoding of signals into spike trains remains a significant challenge.

\subsection{Spiking Neural Networks}
The event-driven mechanisms ensure SNNs consume minimal energy 
, offering a significant advantage for energy and compute-constrained edge devices.
With the introduction of ANN-SNN~\citep{cao2015spiking, stockl2021optimized, bu2022optimized} and STBP~\citep{fang2021incorporating, zhang2021rectified} algorithm, the complexity associated with training high-performance SNNs was significantly reduced.
Based on these advanced learning algorithms, ~\citep{hu2021spiking,zheng2021going} proposed deep residual SNNs and ~\citep{yao2023attention,zhu2022tcja} further contributed multi-dimensional spiking attention mechanisms, achieving commendable performance on ImageNet~\citep{deng2009imagenet} and a variety of neuromorphic datasets~\citep{li2017cifar10, amir2017low}. Subsequently, spiking transformers~\citep{zhou2022spikformer,yao2024spike,yao2023spike} optimized the computation process of spike-based self-attention, further enhancing the performance of SNNs across numerous tasks.
Yet, with the incremental enhancement in performance, there is a notable rise in both the model complexity and memory requirements of these networks, which stands in opposition to the edge-friendly objective.

To further reduce SNNs’ model size, some researchers 
try to increase the network's information capacity. Such as PLIF~\citep{fang2021incorporating} utilizes learnable decay to boost neurons' temporal processing ability, enhancing learning and convergence rates; ~\cite{guo2023ternary} introduced the ternary spike SNN, which increases the network's information capacity at the neuron scale, achieving enhanced performance with smaller model sizes. 
However, these methods still employ full-precision weights and membrane potentials, necessitating higher memory requirements. To address this issue, some researchers~\citep{hu2021quantized,li2022quantization,sulaiman2020weight} have introduced weight quantization into SNNs.
For example, ADMM~\citep{deng2021comprehensive} optimizes a pre-trained full-precision network for low-precision weight quantization. ~\cite{chowdhury2021spatio} utilizes K-means clustering quantization to maintain reasonable accuracy in 5-bit weight SNNs. 
Subsequently, ~\cite{yin2023mint} highlights the significance of quantifying membrane potential in diminishing the memory footprint of SNNs, and quantizes both synaptic weights and membrane potentials, further reducing SNNs' memory requirements. However, with the gradual reduction in bit-width for synaptic weights and membrane potentials, the information representation capability of these networks significantly diminishes, leading to notable performance degradation. Therefore, maintaining network performance and significantly reducing the memory and energy requirements of SNNs remains a challenge.

\section{Preliminaries}
In this section, we present the fundamental principles of threshold-based signal encoding methods. Subsequently, we analyzed traditional binary SNNs' limitations in processing ternary spike trains and introduced the fundamentals and advantages of ternary spike SNNs.
\begin{figure*}[htpb] 
    \centering
  \subfigure[]{
        \label{FigSF1}
 	\includegraphics[scale=0.67]{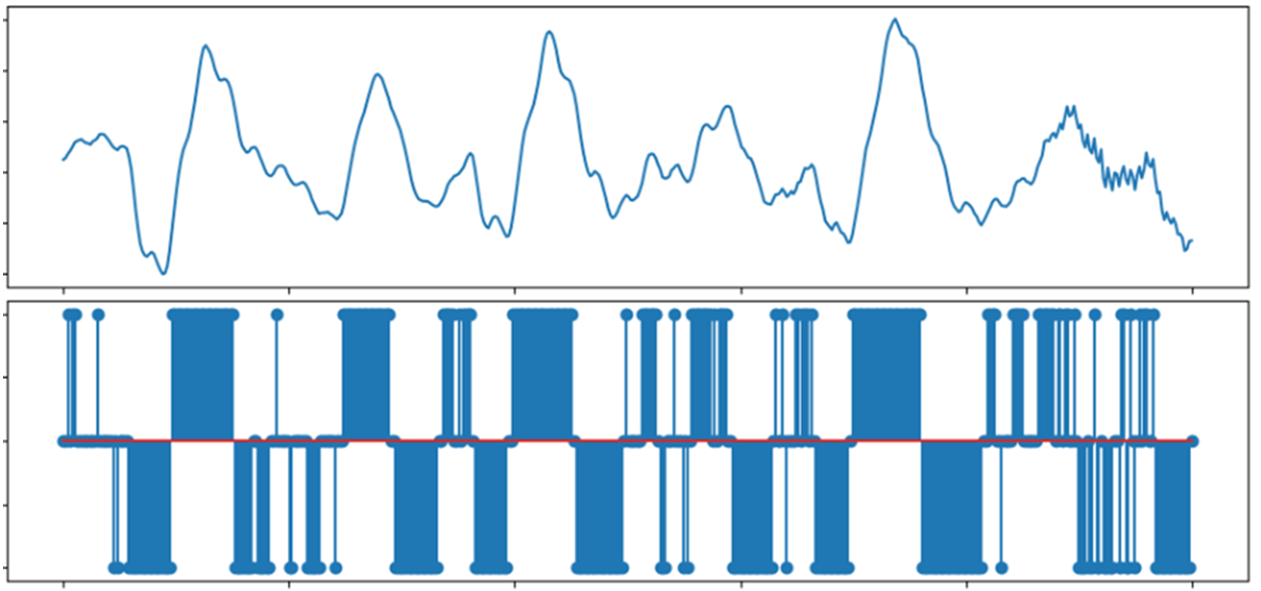}
   }  
 \subfigure[]{
        \label{FigSF2}
 	\includegraphics[scale=0.63]{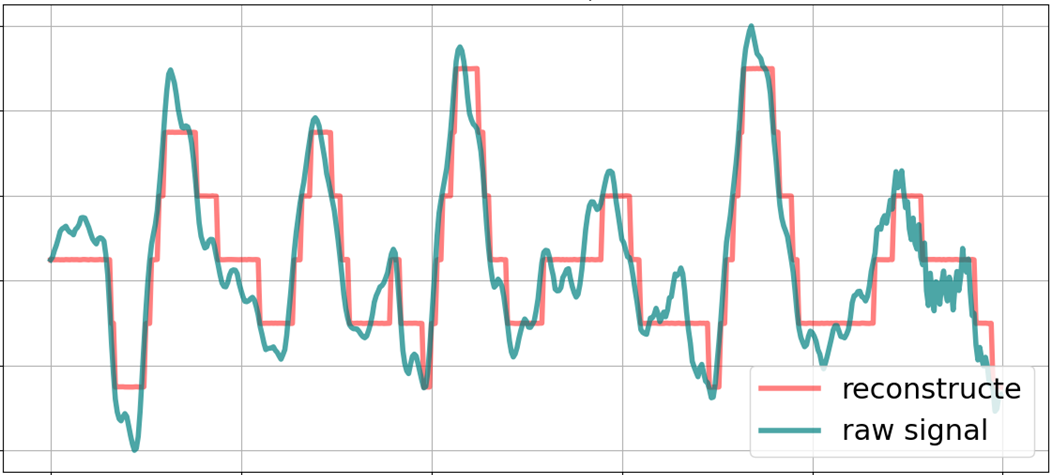}
   }
    \subfigure[]{
        \label{FigSF3}
 	\includegraphics[scale=0.61]{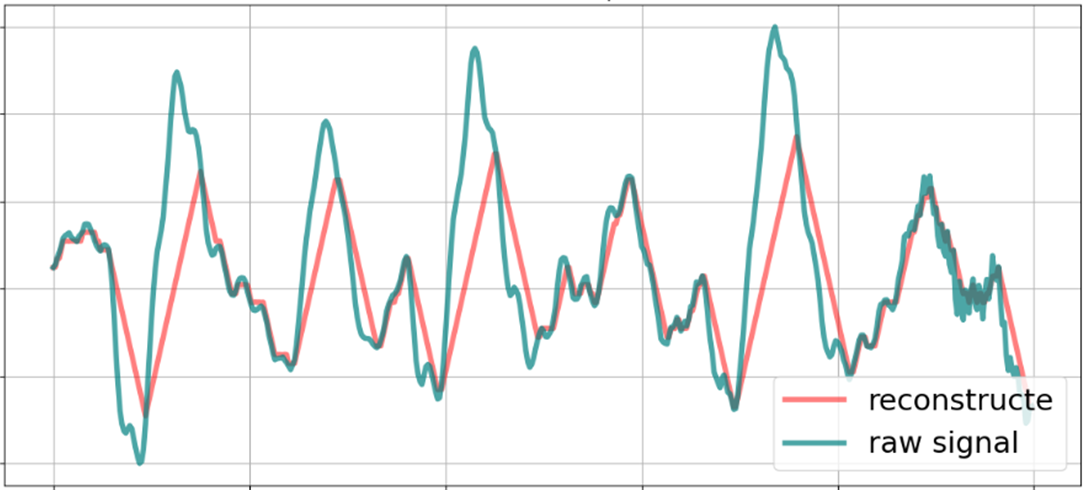}
   }
    \caption{The threshold-based encoding method for raw signals. (a)The threshold-based encoding methods transform raw signals into ternary spike trains consisting of $\left\{-1,0,1\right\}$. (b) A larger threshold may result in notable
reconstruction fluctuations in smooth signal regions. (c) Conversely, a smaller threshold may result in significant peak information loss..}
    \label{picture6}
\end{figure*}
\subsection{ Threshold-based Encoding for Raw Signals}
Although neuromorphic sensors for signals have been developed~\cite{chan2007aer,bartolozzi2018neuromorphic}, their widespread implementation remains limited. Consequently, most SNN-based signal-processing tasks still rely on analyzing analog signals captured by traditional sensors. 
Thus, designing real-time, efficient signal encoding schemes is a major focus. They could transform analog signals from traditional sensors into spike-based signals, significantly reducing information transmission latency and bandwidth requirements.

Within the current research, three threshold-based encoding algorithms stand out: Threshold-based Representation(TBR), Moving Window(MW), and Step-forward(SF) methods. As depicted in Fig.\ref{FigSF1}, the threshold-based encoding methods utilize a fixed threshold to determine the baseline of analog signals, thereby encoding them into ternary spike trains. Among them, SF encoding has achieved significant success in various applications. The SF encoding method calculates the differential of the input signal from a baseline and generates a spike (1 or -1) when the change surpasses a predefined threshold. Notably, the baseline itself is updated according to the spike's polarity. The SF encoding algorithm is described in Algorithm.\ref{alg:SF}.

    \begin{algorithm}
    \caption{Step-forward encoding (SF)}
    \label{alg:SF}
    \begin{algorithmic}[1]
    \State \textbf{Data:} input, threshold
    \State \textbf{Result:} spikes, init
    \State $L \gets \text{length}(input)$
    \State $spikes \gets \text{zeros}(1,L)$
    \State $init, base \gets input(1)$
    \For{$i = 2$ \textbf{to} $L$}
        \If{$input(i) > base + threshold$}
            \State $spikes(i) \gets 1$
            \State $base \gets base + threshold$
        \ElsIf{$input(i) < base - threshold$}
            \State $spikes(i) \gets -1$
            \State $base \gets base - threshold$
        \EndIf
    \EndFor
    \end{algorithmic}
    \end{algorithm}


\subsection{Ternary Spiking Neural network}
Traditional binary spiking neurons cannot directly process and compute negative spikes in those threshold-based encoding methods. This limits the efficient integration between neural signal encoding and backend signal processing models. 
Therefore, to accommodate ternary neural encoding and enhance the information capacity of SNNs, we turn to use ternary Leaky Integrate-and-Fire(TLIF) neurons(\cite{guo2023ternary}). Their neural dynamics can be expressed as follows:

\begin{equation}
u_{i}^{t} = \tau u_{i}^{t-1}Reset\left(U_i^{t-1}\right) + \sum_{j} w_{ij} o_{j}^{t}\label{9}
\end{equation}

\begin{equation}
o_{i}^{t} = 
\begin{cases}
    1, & \text{if } u_{i}^{t} \geq V_{th} \\
    -1, & \text{if } u_{i}^{t} \leq -V_{th} \\
    0, & \text{otherwise}
\end{cases}\label{10}
\end{equation}
where $\tau$ is the constant leaky factor, $u_i^{t+1}$ is the membrane potential of neuron $i$ at the time step $t+1$, and $\sum_{j} w_{ij}^no_j^{t+1}$ denotes
the pre-synaptic inputs for neuron $i$.
When the membrane potential $u_i^{t+1}$ exceeds the firing threshold $V_{th}$, the neuron 
$i$ fires a spike $o_i^{t+1}$ and $u_i^{t+1}$ reset to 0. Eq.~\ref{10} describes the firing function and $Reset\left(u_i^{t+1}\right)=(1 - |o_{i}^{t+1}|)$ is hard reset mechanism.
The ternary spike neuron enhances neuronal information capacity while maintaining the benefits of event-driven processing and full-precision floating-point addition. 

Firstly, we analyze the increased information capacity of ternary spiking neurons from Shannon's Theorem and information entropy. Information entropy, denoted as \( H(X) \), is a metric for the uncertainty or randomness of information in a system, defined by the formula:
\begin{equation}
    H(X) = -\sum_{i=1}^{n} P(x_i) \log_b P(x_i)
\end{equation}
Where \( X \) represents a random variable indicating the state of the system, \( P(x_i) \) is the probability of state \( x_i \), \( n \) is the number of states, and \( b \) is the base of the logarithm, typically 2 for binary systems. For binary SNNs, each neuron can exist in two states (active or inactive). Assuming equal probability for each state (\( P(0) = P(1) = 0.5 \)), the information entropy $H_2 = 1 \text{ bit}$. In contrast, a ternary spike neuron has three possible states and assumes equal probability for each state (\( P(0) = P(1) = P(-1) = \frac{1}{3} \)), yields an information entropy $ H_3 \approx 1.585 \text{ bits}$. Comparatively, the ternary spike neuron exhibits an increased information capacity of approximately 0.585 bits per neuron over the binary spike neuron.

Secondly, ternary spike neurons also exhibit characteristics of event-driven computation and full floating-point addition. As depicted in Eq.\ref{10}, only when the membrane potential exceeds or falls below a threshold value $\pm V_{\text{th}} $(-1, 0, and 1), can ternary spike neurons emit spikes; otherwise, they remain silent. During the accumulation of membrane potentials, owing to the ternary input nature of \( o_j \) (encompassing 0, 1, and -1), the membrane potential update \( u_t \) can be implemented through full floating-point addition, as illustrated in Eq.\ref{9}. Consequently, the ternary spike neuron model fits ternary neural encoding methods, while retaining the advantages of binary spike SNNs and enhancing the information capacity of SNNs.

\section{Method}
In this section, we introduce the TAE method and QT-SNN to address the bottlenecks mentioned in SNN-based intelligent signal processing tasks.
Specifically, the TAE method can encode analog signals into ternary spike trains with 
adaptive threshold, exhibiting greater sparsity, and better performance. Meanwhile, QT-SNN tailors to the TAE method, and significantly reduces the energy consumption and memory requirements of backend intelligent signal processing models. Next, we will introduce the details of the TAE method and QT-SNN separately.

\subsection{Threshold-Adaptive Encoding }
As previously mentioned, those threshold-based signal encoding methods efficiently encode analog signals into ternary spike trains, significantly reducing the bandwidth required for information transmission.
Nevertheless, their efficacy largely hinges on the initial $threshold$ setting. For example, in the SF algorithm.\ref{alg:SF}, $threshold$ is typically set based on extensive experimentation, rather than being dynamically adjusted in response to ongoing signal changes. This static approach may cause some problems. As illustrated in Figs.\ref{FigSF2} and \ref{FigSF3}, a larger threshold can cause notable reconstruction fluctuations in smoother signal regions, and a smaller threshold may result in significant peak information loss. Consequently, a fixed threshold struggles to accommodate the intricacies of time-varying signals fully.
To address this limitation, ~\cite{de2021population} have introduced a group-based SF encoding method. This approach utilizes multiple thresholds to more effectively extract information from both smooth and peak regions of the signal. While this strategy reduces information loss, it also increases computational demands. Therefore, the development of an adaptive threshold mechanism, which dynamically adjusts to the current signal variations, remains a critical challenge in optimizing SF encoding for complex, time-varying signals.

To better address the loss of information due to fixed thresholds, we propose a threshold-adaptive encoding(TAE) method. This scheme adapts the threshold dynamically in response to the current variations in signals, significantly reducing the information loss of encoding in the smoothing and steep phase. The TAE method is shown as Algorithm.\ref{alg:ASF}.
\begin{algorithm}
\caption{Threshold-Adaptive Encoding(TAE)}
\label{alg:ASF}
\begin{algorithmic}[1]
\State \textbf{Data:} input, threshold, $a$
\State \textbf{Result:} spikes, init
\State $L \gets \text{length}(input)$
\State $spikes \gets \text{zeros}(1, L)$
\State $init, base \gets input(1)$
\For{$i = 2$ \textbf{to} $L$}
    \If{$input(i) \geq base + threshold$}
        \State $spikes(i) \gets 1$
        \State $base \gets base + threshold$
        \State $threshold \gets threshold \times a$
    \ElsIf{$input(i) \leq base - threshold$}
        \State $spikes(i) \gets -1$
        \State $base \gets base - threshold$
        \State $threshold \gets threshold \times a$
    \Else
        \State $threshold \gets threshold / a$
    \EndIf
\EndFor

\end{algorithmic}
\end{algorithm}
\begin{figure}[htpb] 
    \centering
  \subfigure[]{
        \label{Fig.sf}
 	\includegraphics[scale=0.31]{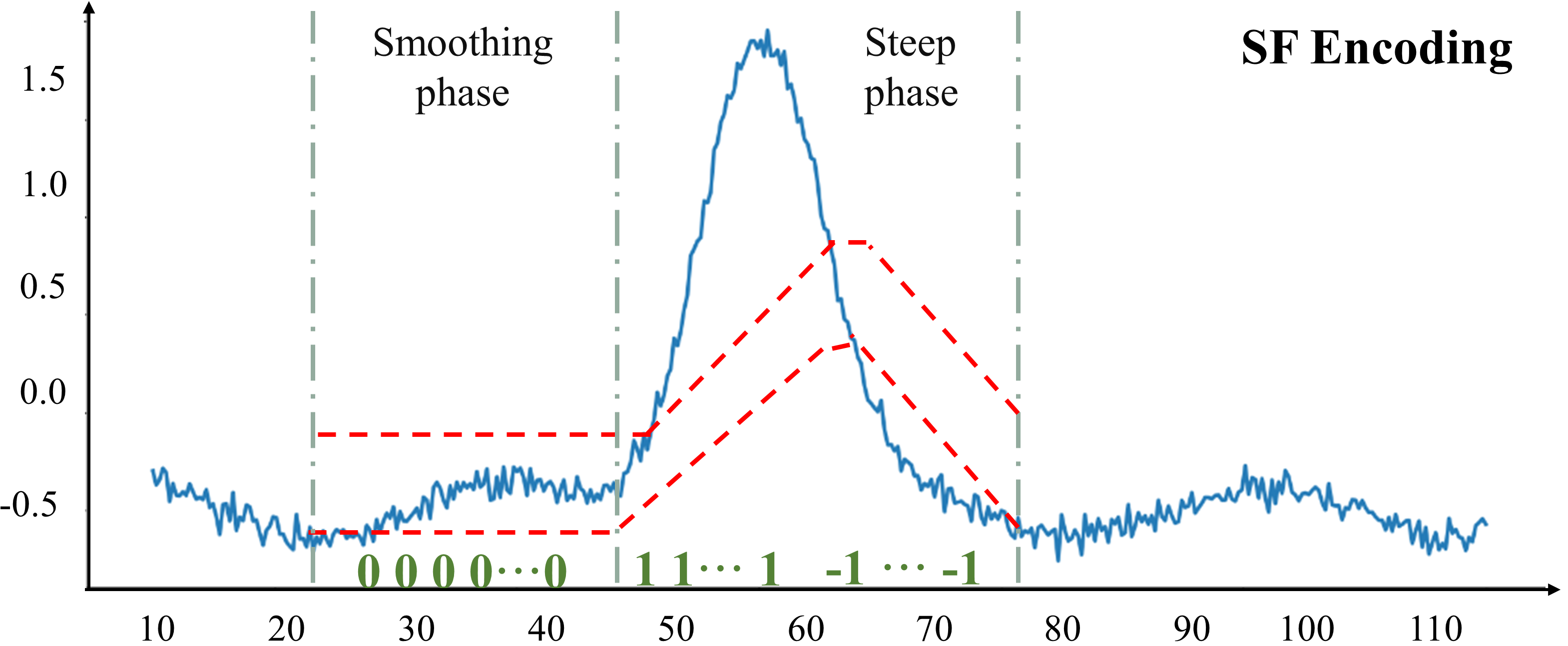}
   }  
 \subfigure[]{
        \label{Fig.asf}
 	\includegraphics[scale=0.31]{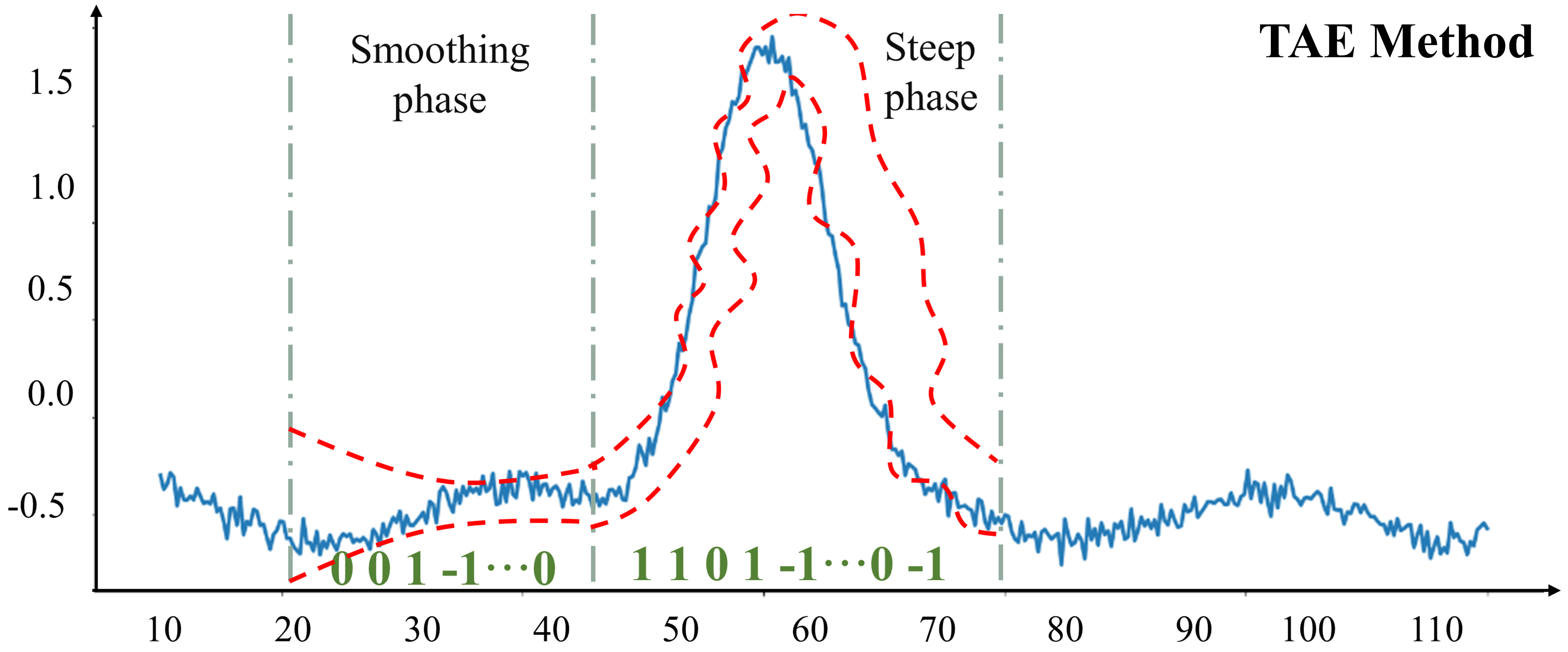}
   }
    \caption{The comparison between SF and TAE methods. (a) In SF coding, the fixed threshold results in a linear increase and decrease of the $base$ (indicated in red).  (b) The TAE method employs an adaptive thresholding technique, allowing the base to conform to any given curve (red cave).}
    \label{picture7}
\end{figure}
In our TAE method, $Input$ refers to the initial signal input, $ Threshold$ denotes the initial threshold value, whereas $a$ represents the hyperparameter for threshold adaptation, which is typically preset to a value of 1.1. The essence of this algorithm lies in its capability to autonomously modulate the threshold size in response to the current dynamic state of the signal. As demonstrated in Fig.\ref{Fig.asf}, when the signal is in a smoothing phase, TAE gradually decreases the threshold through $a$ until the $base \pm threshold$ can detect changes in the smoothing signal. Conversely, during peak phases, it increases the threshold by $a$ to accommodate the signal's rapid fluctuations.
This approach can effectively address the issue of losing smooth information and prominent peaks in threshold-fixed encoding as shown in Fig.\ref{Fig.sf}, retaining all the advantages of threshold-based neural coding methods. Furthermore, TAE minimizes the occurrence of continuous spikes states in the encoding output, such as $\left\{-1, -1, …, -1\right\}$ and $\left\{1, 1, …, 1\right\}$. This results in a lower spiking firing frequency for information transmission. we will demonstrate this point in the experiments of Section 4.1.
\begin{figure*}[htpb] 
    \centering
  \subfigure[]{
        \label{Fig.sub.q1}
 	\includegraphics[scale=0.62]{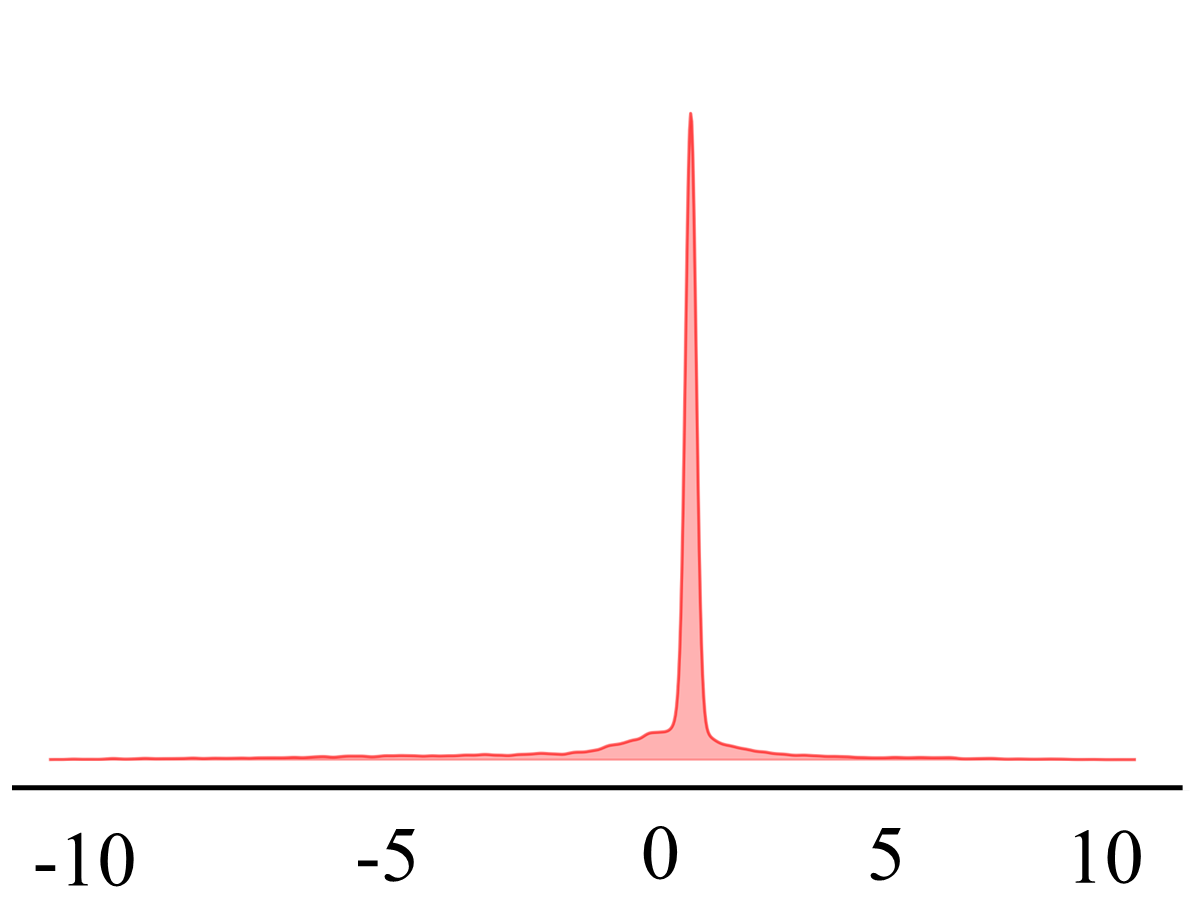}
   }  
   \subfigure[]{
        \label{Fig.sub.q2}
 	\includegraphics[scale=0.62]{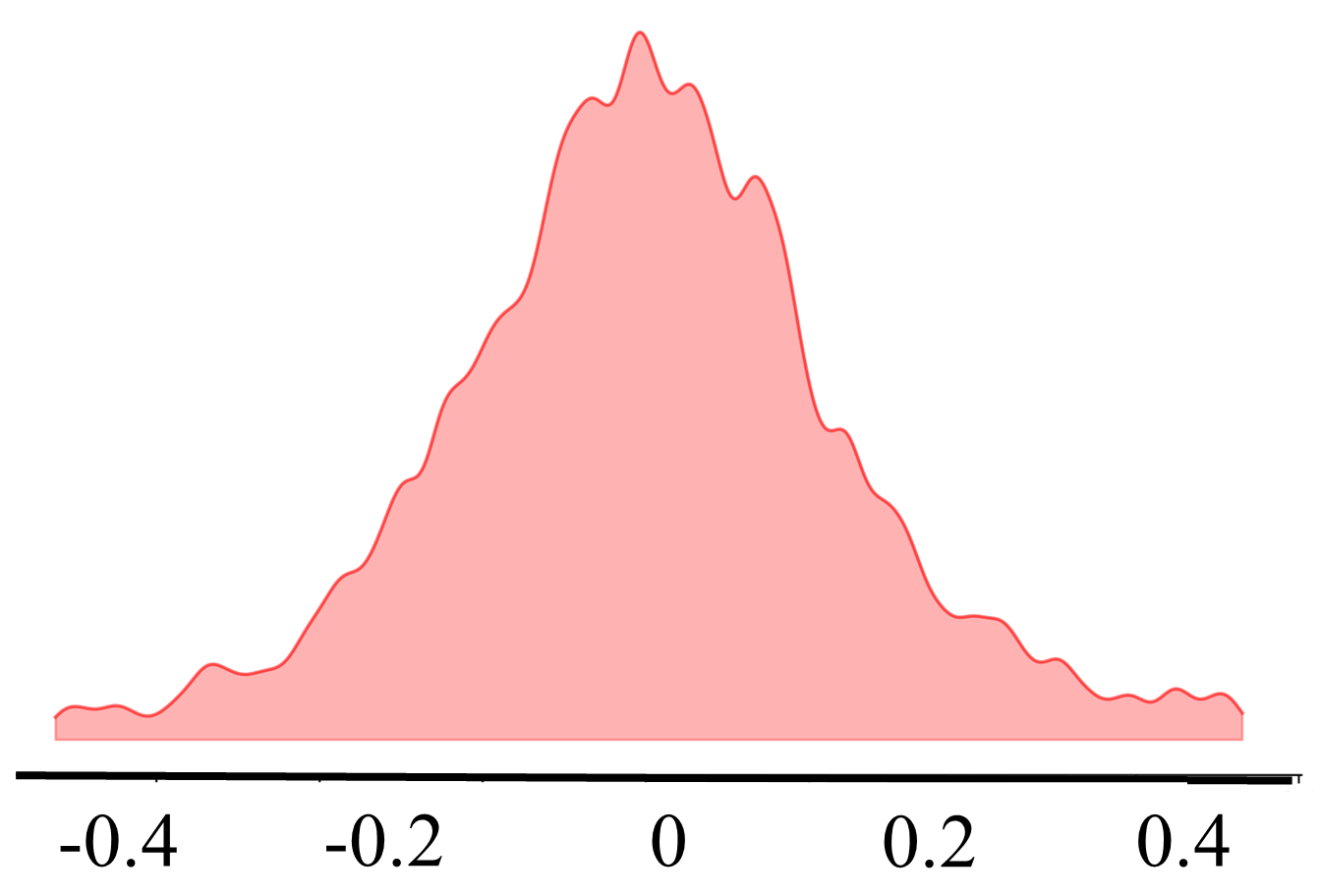}
   } 
   \subfigure[]{
        \label{Fig.sub.q3}
 	\includegraphics[scale=0.62]{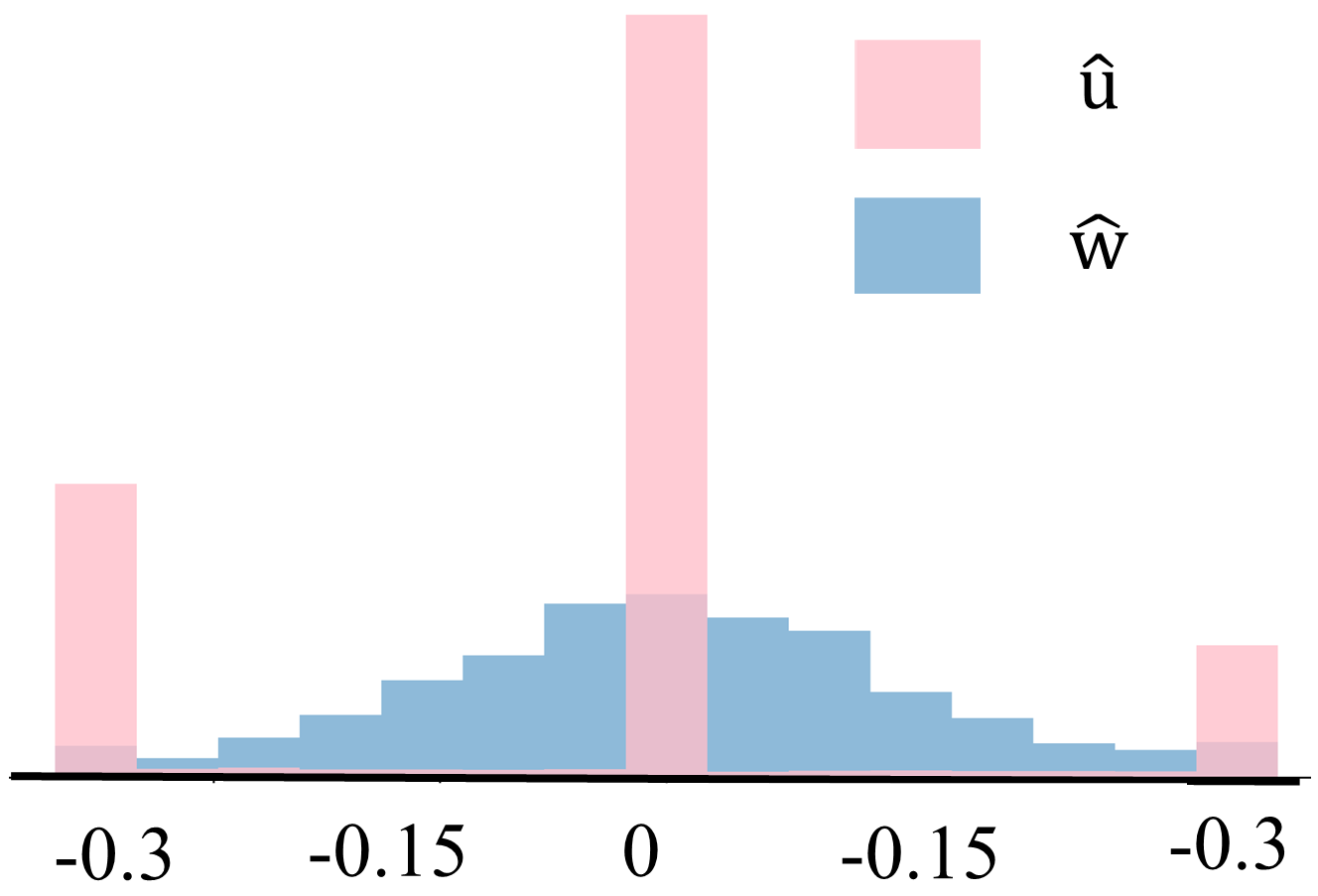}
   } 

    \caption{The distribution of weights and membrane potential. (a) In a full-precision ternary spike SNN, the membrane potential and weights follow normal distributions with a mean of zero and differing standard deviations. (b) The MINT method quantizes the membrane potential and weights into 4 bits using nonlinear mapping function \(\tanh\), disrupting the normal distribution characteristics of the membrane potential.}
    \label{picture6}
\end{figure*}
\subsection{Dual-scaling Factor Quantization Ternary Spiking Neural Networks}
Despite ternary spike SNNs capitalize on the energy-efficiency benefits of event-driven processing and can directly process ternary spikes from our TAE method. The membrane potentials and synaptic weights in them remain represented as full-precision floating-point values, as illustrated in Eq.\ref{9}. This presents two significant issues: firstly, the deployment of models also requires substantial memory footprints. secondly, operations with full-precision floating values are still not energy-efficient. To address these challenges, many researchers~\cite{castagnetti2023trainable, hwang2024one} have applied quantization techniques to the synaptic weights to achieve reduced memory consumption and enhanced energy efficiency. Suppose weights are represented by $ w\in \mathbb{R}^{C_{\text{out}} \times C_{\text{in}}}$, where $C_{\text{out}}$ and $C_{\text{in}}$ are the number of output and input channels respectively. The quantization of the weights is denoted as:

\begin{equation}
\widehat{w} = \Pi_{\mathcal{Q}_{\alpha, b}}[w, \alpha]
\end{equation}

where $\alpha$ is the clipping threshold and the clipping function $\Pi_{\alpha}$
clips weights into the interval $[-\alpha, \alpha]$. After clipping, $w$ is projected by $\Pi(\cdot)$ onto the quantization levels $\widehat{w}$. We define 
$\mathcal{Q}(\alpha, b)$ for a set of quantization levels, where $b$ is the bit-width.
For uniform quantization, the quantization levels are defined as

\begin{equation}
\mathcal{Q}(\alpha, b) = \alpha \times \left\{0, \pm\frac{1}{2^{b-1}-1}, \pm\frac{2}{2^{b-1}-1}, \ldots, \pm1\right\}. 
\label{eq.qua}
\end{equation}

For every floating-point number, uniform quantization maps it to a $b$-bit fixed-point representation in $\mathcal{Q}_u(\alpha, b)$. $\alpha$ is stored separately as a full-precision floating-point in $\mathcal{Q}(\alpha, b)$.

However, few works focus on the unique membrane potential in SNNs. To concurrently quantize membrane potential and synaptic weights within SNNs, further reducing the memory requirements of ternary spike SNNs, we initiate the process by straightforwardly applying Eq.\ref{eq.qua}(UQ) to Eq.~\ref{9}, employing separate full-precision scaling factors $\alpha_1$, $\alpha_2$, and $\alpha_3$ for each integer value. At this stage, the updated equation for membrane voltage can be presented as:

\begin{equation}
\alpha_1 \widehat{u_{i}^{t}} = \alpha_2 \tau \widehat{u_{i}^{t-1}}Reset\left(U_i^{t-1}\right) + \sum_{j} \alpha_3 \widehat{w_{ij}} o_{j}^{t}\label{11}
\end{equation}


However, directly applying Eq.\ref{11} for training and inference introduces several MAC operations during the accumulation of membrane potential, compromising the energy efficiency of SNNs. ~\cite{yin2023mint} addressed this by leveraging a nonlinear mapping function to map both \(u\) and \(w\) to the range \([-1,1]\), thereby ensuring \(a_1 = a_2 = a_3\). This allowed \(a_1\) to be integrated into the threshold for spike emission. Nevertheless, as shown in Fig.\ref{Fig.sub.q1} and Fig.\ref{Fig.sub.q2}, the distributions of \(u\) and \(w\) indicates a significant discrepancy. Consequently, to better maintain the distribution of membrane potential and synaptic weights, we introduce a dual-scale factor quantization strategy. The membrane potential update formula is presented as follows:

\begin{equation}
 \alpha_1 \widehat{u_{i}^{t}} =  \alpha_1 \tau\widehat{ u_{i}^{t-1}}Reset\left(U_i^{t-1}\right) + \sum_{j} \alpha_3 \widehat{w_{ij}} o_{j}^{t}\label{12}
\end{equation}

In Eq.\ref{12}, We assume that the membrane potentials across different time steps follow similar normal distributions, thus allowing the use of a uniform scaling factor \(a_1\). Conversely, synaptic weights are scaled using a distinct factor \(a_3\). To further enhance the energy efficiency of our QT-SNN in inference, we amalgamate Eq.\ref{11} and Eq.\ref{10}, integrating $\alpha_1$ and $\alpha_3$ into the spike emission process. The revised formulation for membrane potential update and spike emission can be described as :

\begin{equation}
 \widehat{u_{i}^{t}} =   \tau \widehat{u_{i}^{t-1}}Reset\left(U_i^{t-1}\right) + \sum_{j} \widehat{w_{ij}} o_{j}^{t}\label{13}
\end{equation}

\begin{equation}
o_{i}^{t} = 
\begin{cases}
    \frac{\alpha_3}{\alpha_1}, & \text{if } u_{i}^{t} \geq \lceil V_{th} / \alpha_1 \rceil \\
    -\frac{\alpha_3}{\alpha_1}, & \text{if } u_{i}^{t} \leq \lceil -V_{th} / \alpha_1 \rceil \\
    0 , & \text{otherwise}
\end{cases}\label{14}
\end{equation}

During the training phase, as demonstrated in Eq.\ref{13}, the update of the membrane potential avoids MAC operations. However, during the spike emission process, the intensity of spikes is scaled by the ratio \(a_3/a_1\), and the threshold is based on \(V_{\text{th}}/\alpha_1\), both of which are parameters subject to learning. To ensure quantization performance and further reduce the energy consumption during the inference phase, we set \(V_{\text{th}}/\alpha_1\) as a full-precision learnable parameter. The membrane decay factor \(\tau\) is set to \(2^{-1}\), and \(a_3/a_1\) is also defined as a learnable parameter in the form of powers of 2. This configuration allows for the utilization of bit-shift operations for both training and inference phases, with the specific bit-shift formula presented as follows:

\begin{equation}
2^{x}r = 
\begin{cases}
    r >> x  , & \text{if } x > 0 \\
    r << x  , & \text{if } x < 0 \\
    r  , & \text{otherwise}
\end{cases}\label{15}
\end{equation}

Subsequently, during the inference phase, the threshold can be calculated using \(\lceil V_{\text{th}}/\alpha_1 \rceil\), where \(\lceil * \rceil\) denotes the ceiling operation; \(a_3/a_1\) can be reintegrated into the membrane voltage accumulation process through bit-shift operations on \(w\), thereby ensuring that spikes are transmitted as \{-1, 0, 1\} throughout the inference process. The specifics of this inference process are described in Algorithm.\ref{alg:INF}.

\begin{algorithm}
\caption{Inference path}
\label{alg:INF}
\begin{algorithmic}[1]
\State \textbf{Data:} 
\State Input spikes $o^t_i$ to the layer i at time t
\State integer weights $w_i$ of the layer i
\State leakage factor $\tau \gets 0.5$
\State integer membrane potential $u^{t-1}_i$ at time t-1
\State integer exponent $k \gets log_2 \frac{\alpha_3}{\alpha_1}$
\State integer thresholds $\theta$ of value 
\State \textbf{Result:} 
\State Output spikes $o^t_i$ to the layer i at time t
\State integer membrane potential $u^t_i$ at time t

\State $X_i^t = \sum_{j} w_{ij} o_{j}^{t}$
\State $H^t_i = X_i^t >> k + u^t_i >> 1$
\If{$H^t_i \geq \theta$}
    \State $o^t_i \gets 1$
    \State $u^t_i \gets 0$
\ElsIf{$H^t_i \leq - \theta$}
    \State $o^t_i \gets 1 $
    \State $u^t_i \gets 0 $
\Else
    \State $o^t_i \gets 0 $
    \State $u^t_i \gets H^t_i$
\EndIf
\end{algorithmic}
\end{algorithm}

Our QT-SNN eliminates MAC operations during the inference process, maximally preserving the energy efficiency advantage of SNNs. Additionally, it quantizes both weights and membrane potentials to lower bit-width,
reducing the memory required to deploy the model on hardware. 
During the training phase, treating \(a_1/a_3\) as a learnable spike value effectively enhances the performance of QT-SNN. we will validate these points in the experimental section. Therefore, employing QT-SNN as the backend model for intelligent signal processing tasks markedly improves performance while simultaneously reducing model deployment memory footprints.

\begin{figure*}[htpb] 
    \centering
   \subfigure[]{
        \label{Fig.SF2}
 	\includegraphics[scale=0.4]{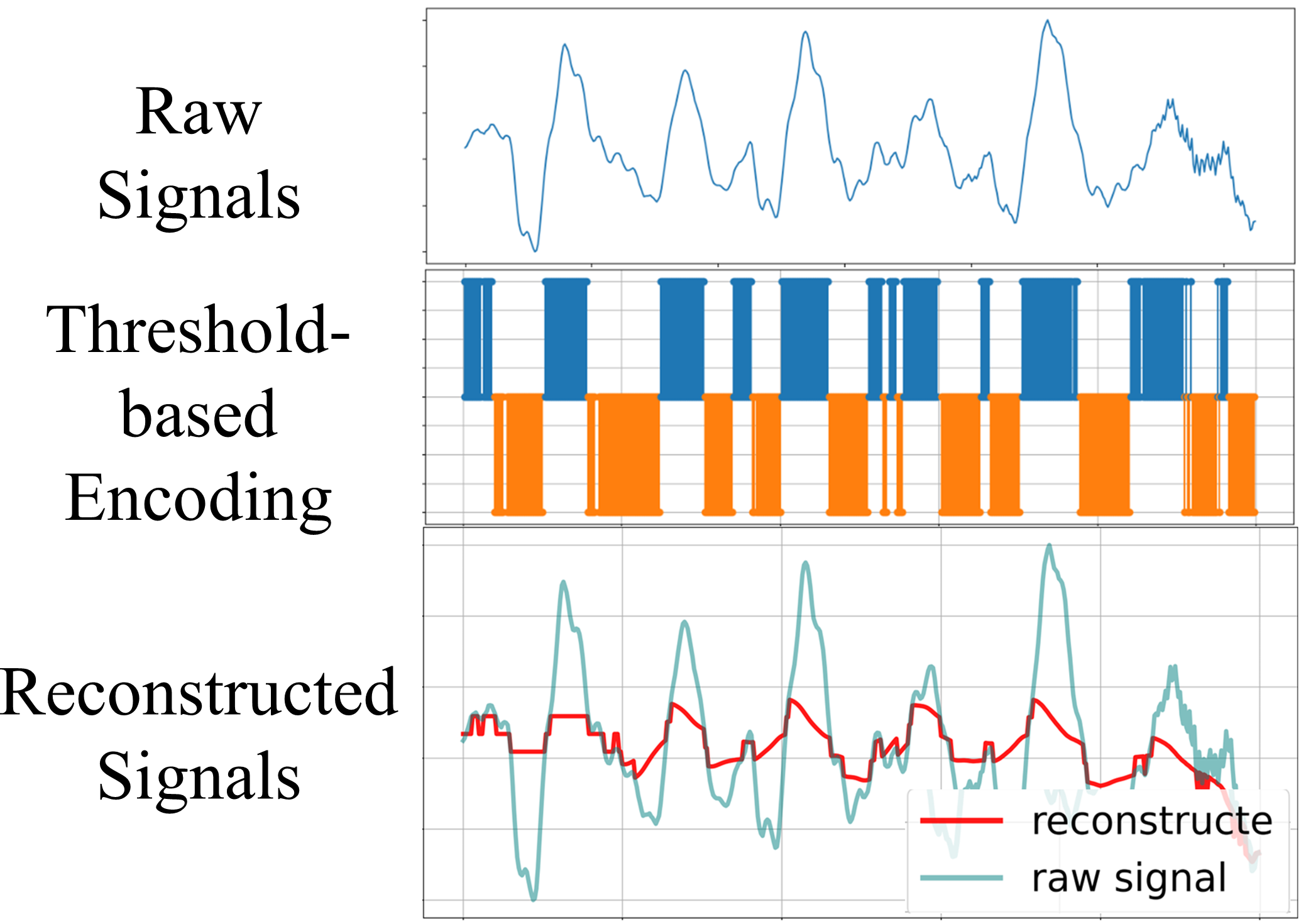}
    }
  \subfigure[]{
        \label{Fig.sub.SF1}
 	\includegraphics[scale=0.39]{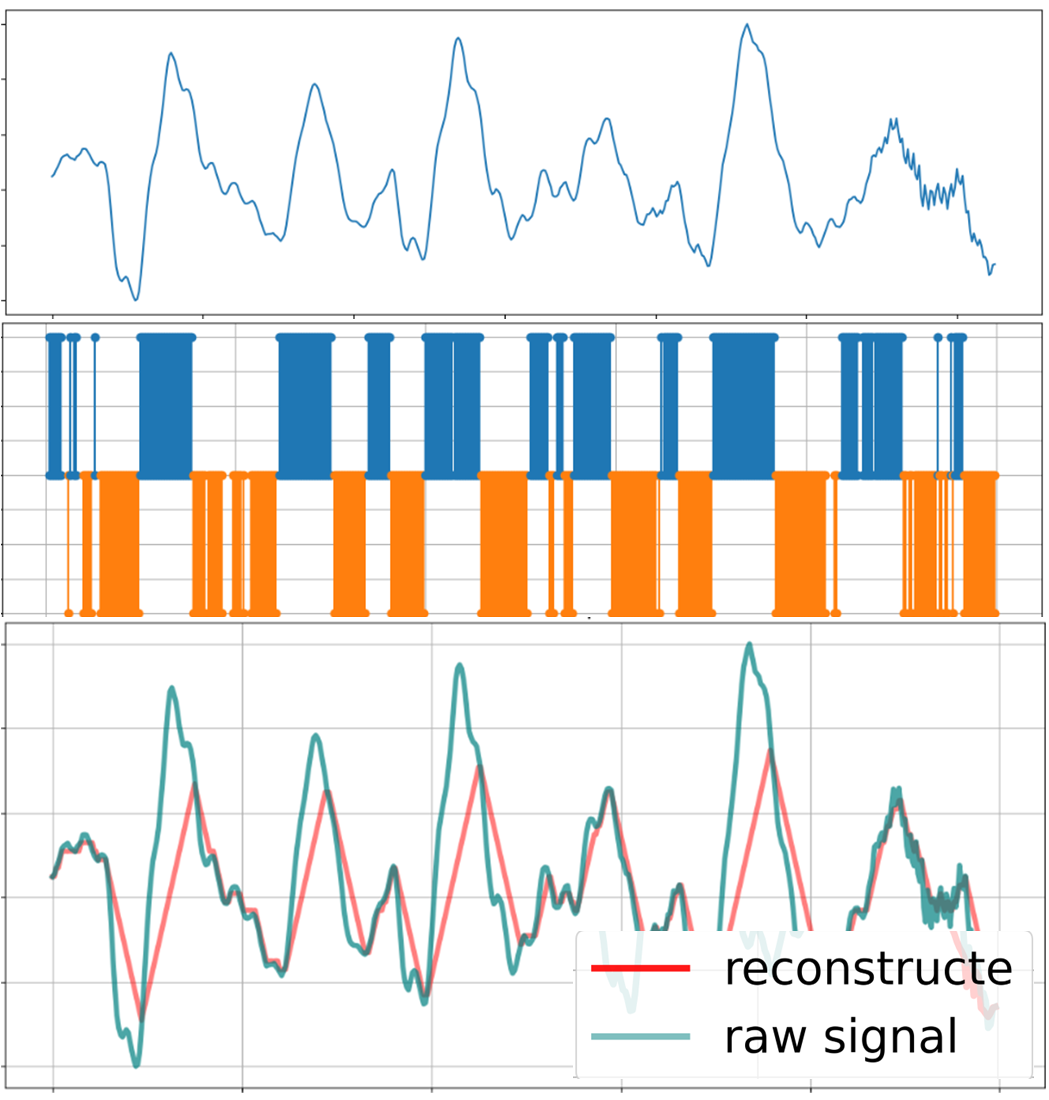}
   }  
 \subfigure[]{
        \label{Fig.SF2}
 	\includegraphics[scale=0.39]{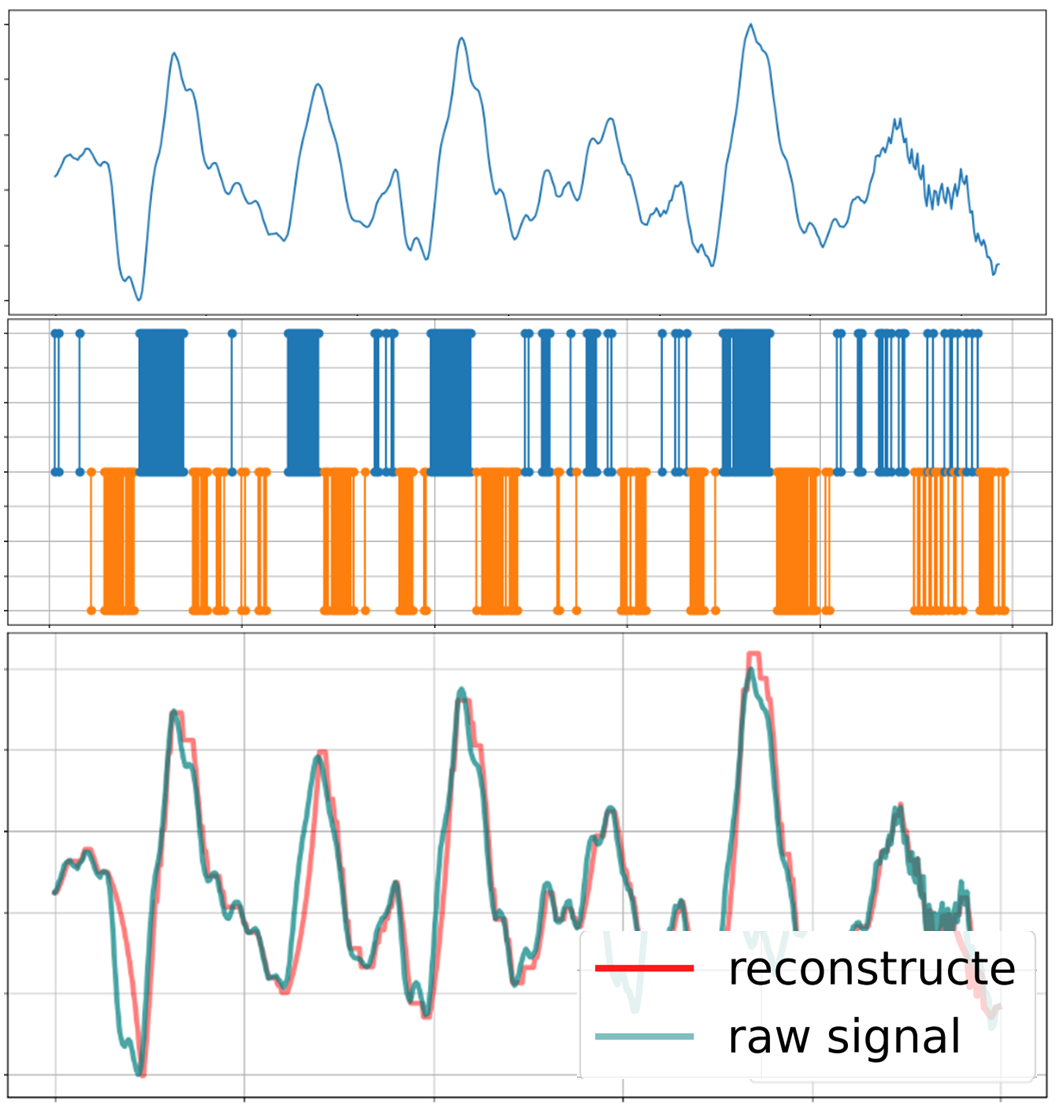}
   }
   \subfigure[]{
        \label{Fig.SF2}
 	\includegraphics[scale=0.29]{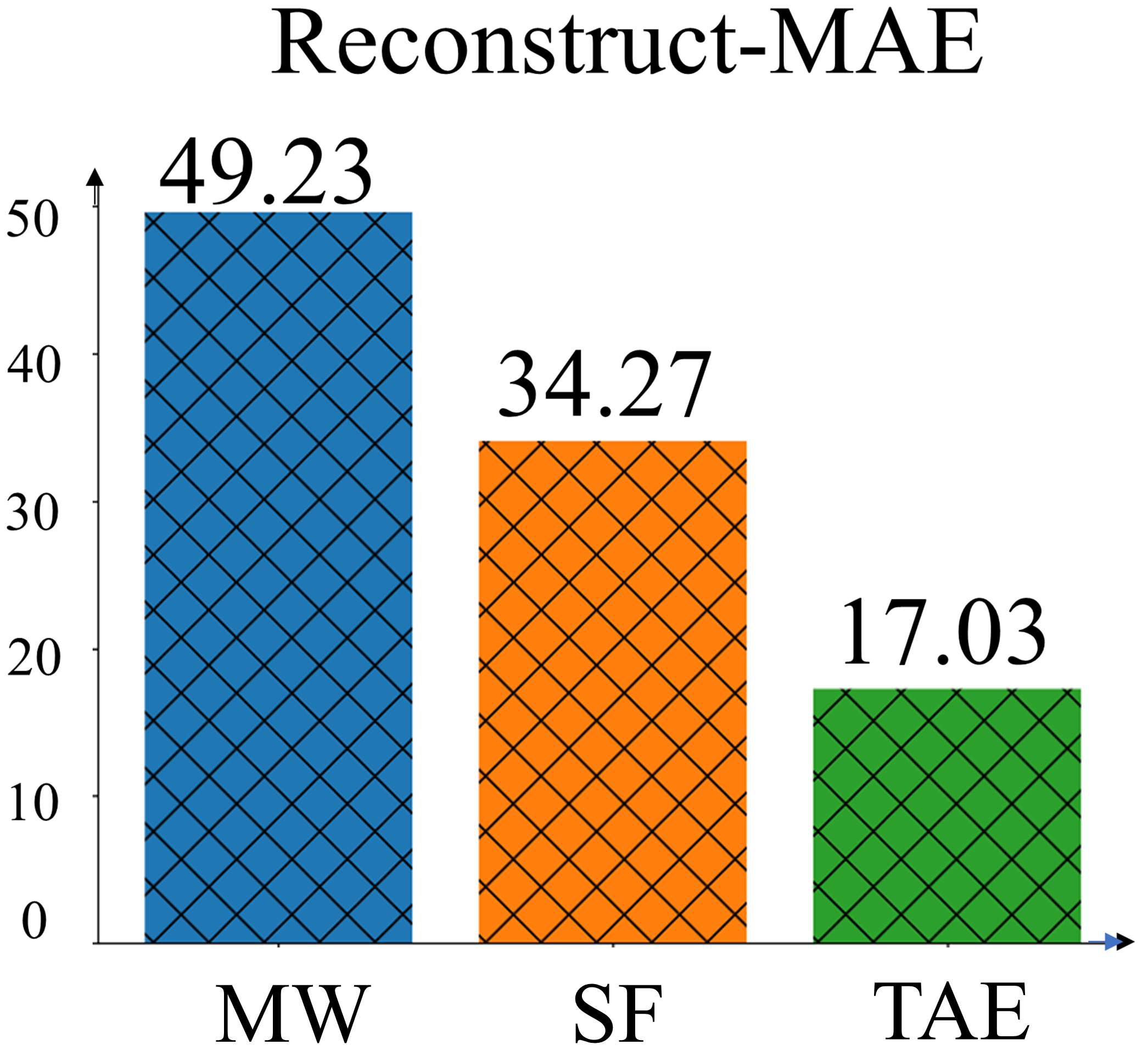}
   }
   \subfigure[]{
        \label{Fig.SF2}
 	\includegraphics[scale=0.38]{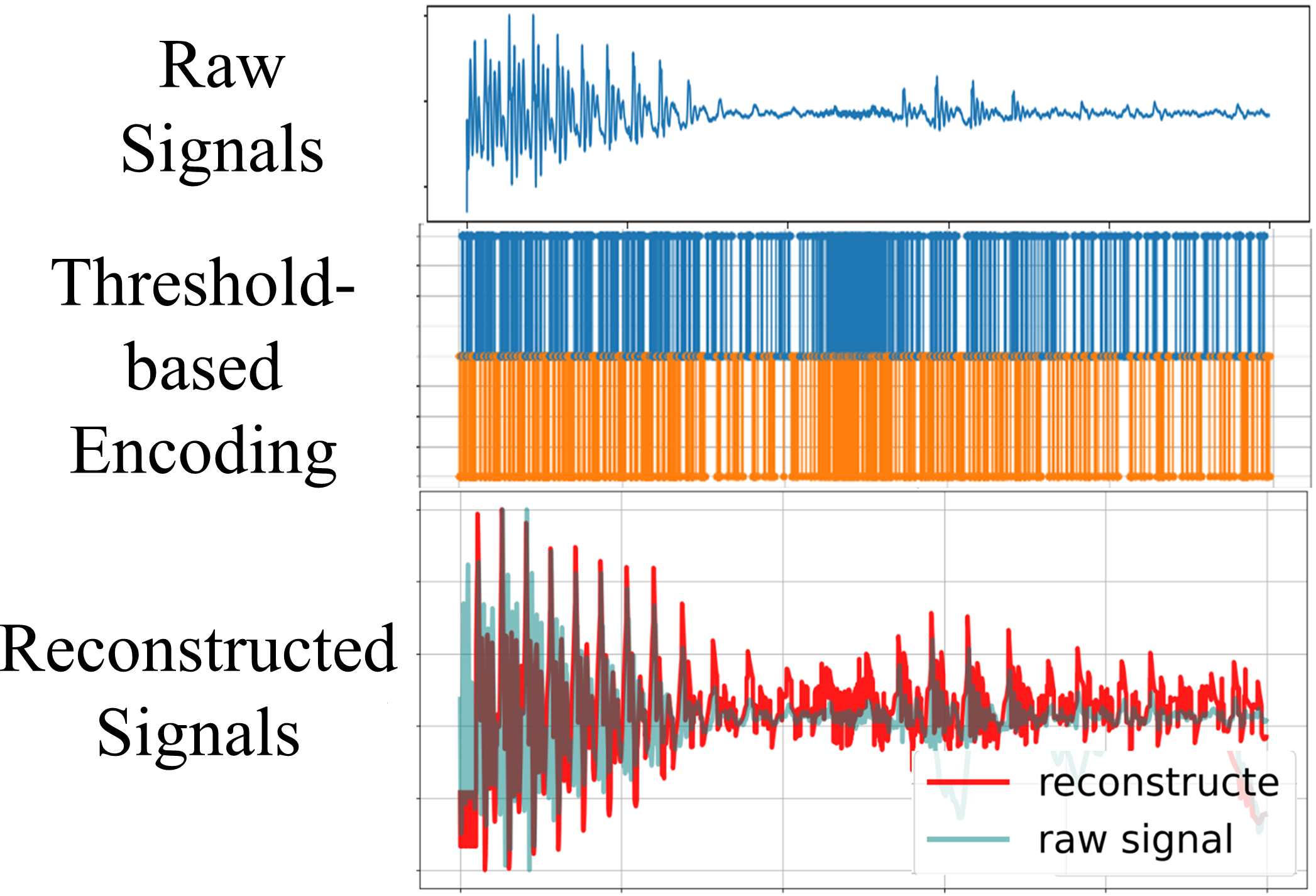}
   }
   \subfigure[]{
        \label{Fig.SF2}
 	\includegraphics[scale=0.39]{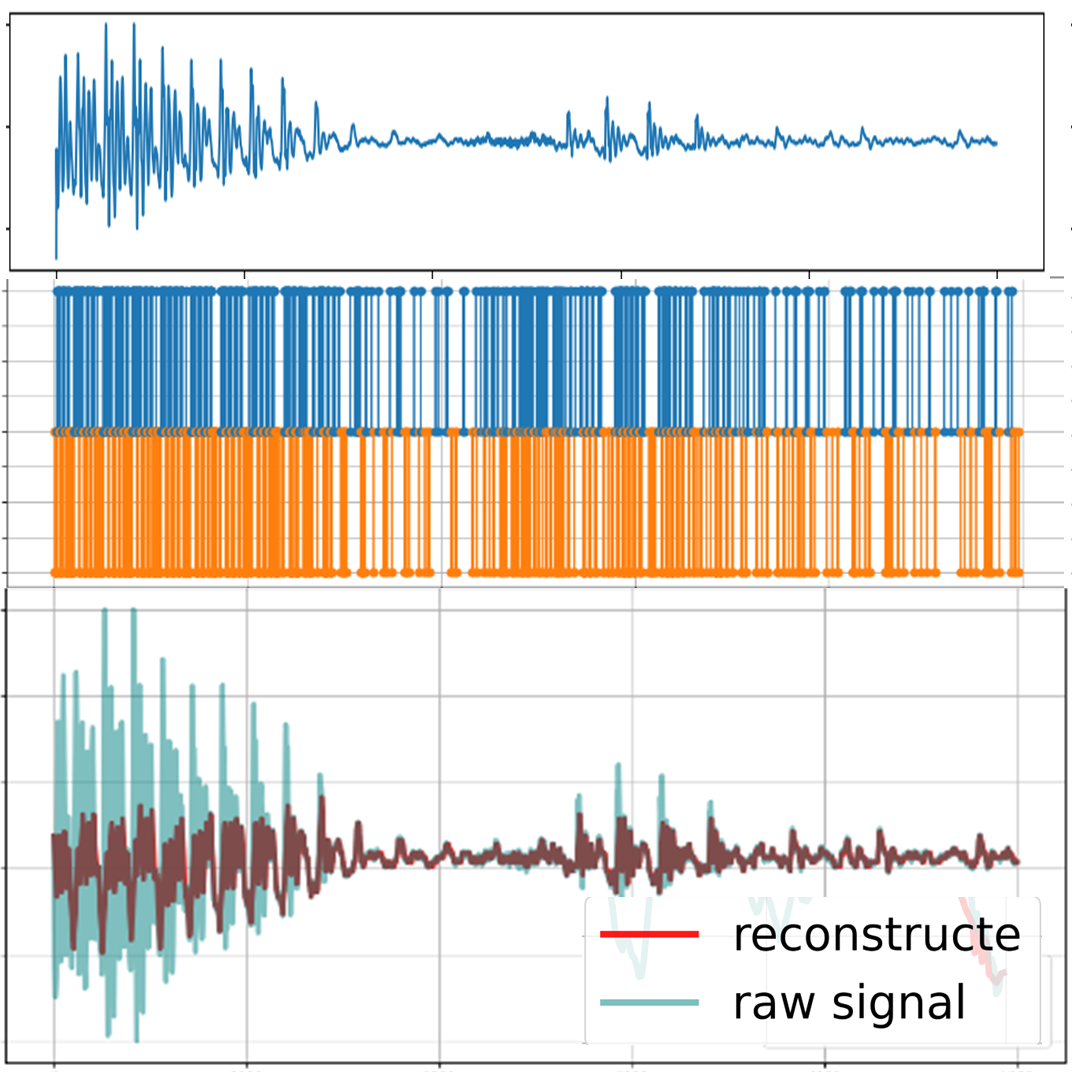}
   }
   \subfigure[]{
        \label{Fig.SF2}
 	\includegraphics[scale=0.39]{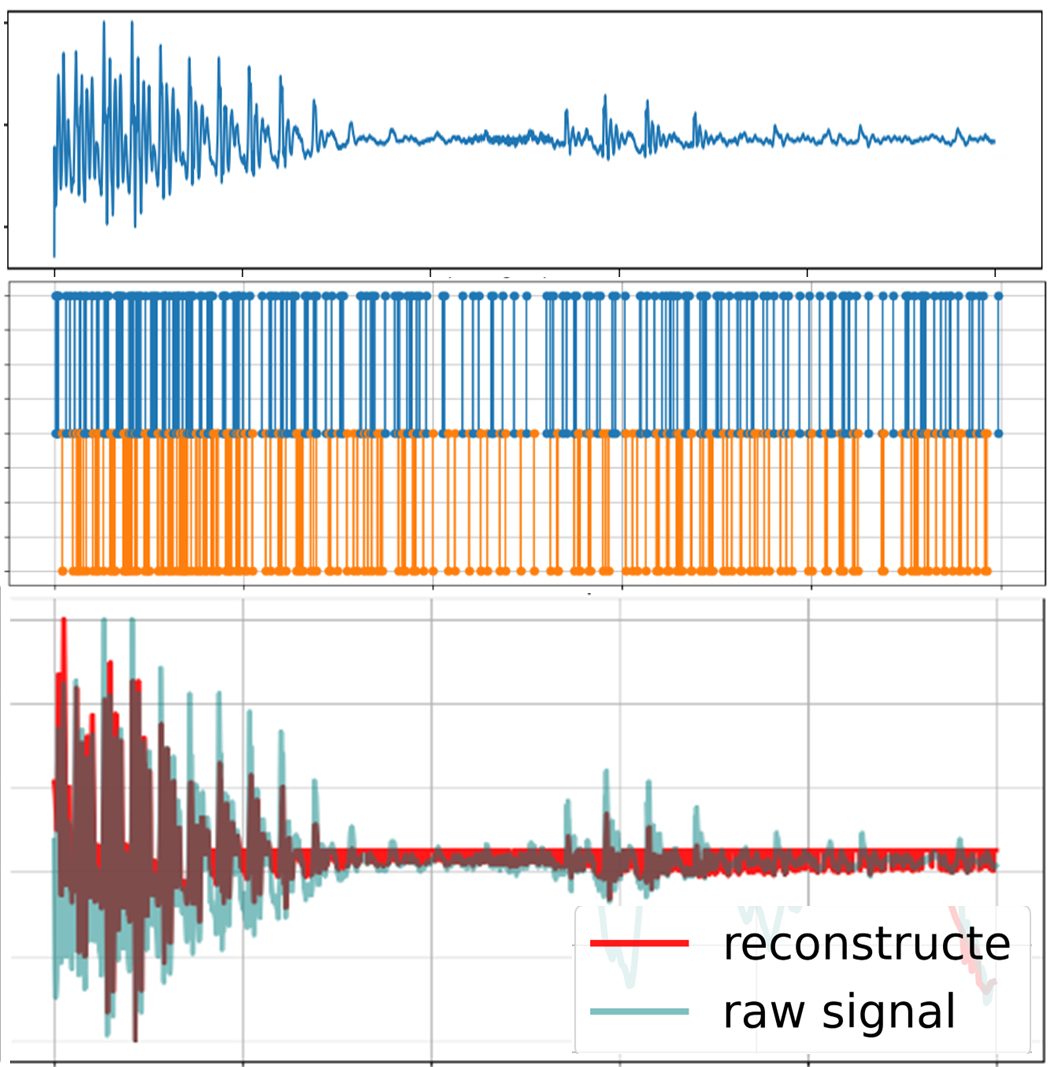}
   }
   \subfigure[]{
        \label{Fig.SF2}
 	\includegraphics[scale=0.29]{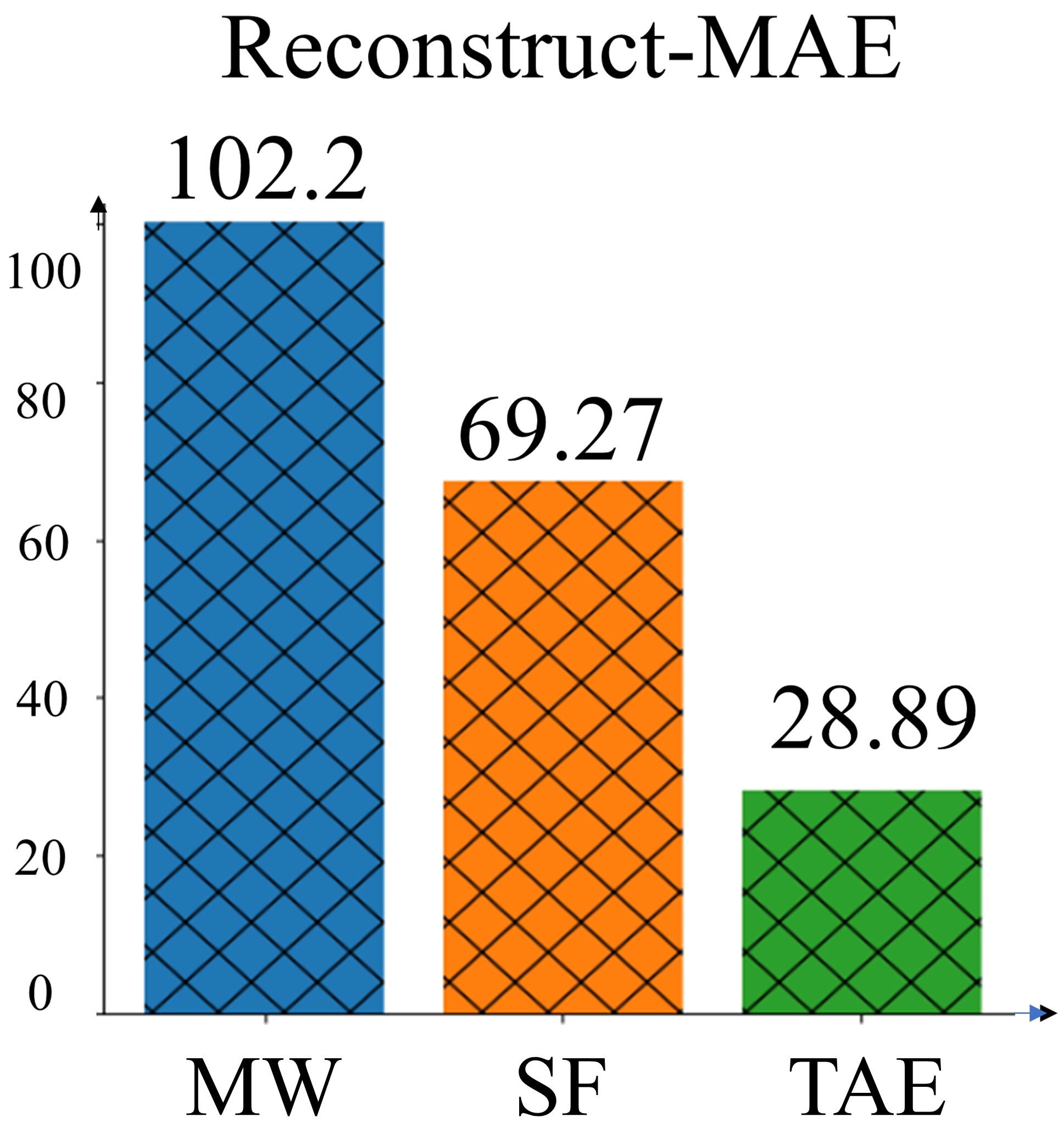}
   }
    \caption{A comparative performance analysis of different encoding methods on radar and GSC datasets. (a-c) depict the encoding and reconstruction capabilities of MW, SF, and our TAE method for a radar signal segment. (e-g) illustrate these encoding strategies applied to a segment of speech signal. (d) and (h) include statistical analysis of the reconstruction-MAE for the overall datasets.}
    \label{picture6}
\end{figure*}

\section{Experiment}

In this section, we systematically validate the performance advantages of our TAE method and QT-SNN through comparative experiments. For TAE method, we assess its spike firing rate and the mean absolute error (MAE) of reconstruction against two other neuromorphic encodings across radar~\citep{lopez2022time} and speech datasets. Regarding QT-SNN, we examine its performance enhancement over other quantized SNNs on the CIFAR10 datasets. Subsequently, we verify the performance and memory footprints of the proposed ternary spike-based neuromorphic signal processing system on the GSC and EEG datasets. Finally, we analyze the energy efficiency benefits of our system compared to ANNs and other SNNs using DC encoding methods.

\subsection{Datasets}
\textbf{GSC Dataset:} The Google Speech Commands(GSC) dataset includes 30 short commands for Version 1 (V1) and 35 for Version 2 (V2), recorded by 1,881 and 2,618 speakers, respectively. 
To make a fair comparison, our experiments are conducted on the 12-class classification and 35-class classification tasks as previous SNN models~\cite{yilmaz2020deep,yang2022deep, orchard2021efficient}. While 12-class classification recognizes 12 classes, that include 10 commands: “yes”, “no”, “up”, “down”, “left”, “right”, “on”, “off”, “stop” “go”, and two additional classes: silence, and an unknown class. The unknown class covers the remaining 20 (25) speech commands in the set of 30 (35). The silence class accounting for about 10 $\%$ of the total dataset is generated by splicing the noise files in the dataset. Finally,
GSC-V1 is split into 56588 training, 7743 validation, and 7835 test utterances, and GSC-V2 is divided into 92843 training, 11003 validation, and 12005 test utterances.

\textbf{KUL dataset:} EEG recordings were obtained from 16 individuals with normal hearing, engaged in the task of concentrating on a specific speaker among two. The auditory stimuli consisted of four tales in Dutch, narrated by three male Flemish speakers. To maintain uniform perceived loudness, the intensities of the stimuli were normalized using root mean square (RMS) and presented dichotically (one speaker per ear) or via head-related transfer function filtering, creating an auditory illusion of speech originating from 90° relative to the listener's position. The experiment comprised eight sessions, each lasting six minutes, with the order of conditions randomized for each participant. The EEG data were collected using a 64-channel BioSemi ActiveTwo system at a sampling rate of 8,192 Hz, within a sound-controlled setting. Detailed information on the expanded KUL dataset can be found in the referenced literature~\citep{das2019auditory,vandecappelle2021eeg}.

\textbf{DTU dataset:} This dataset comprises EEG recordings from 18 participants with normal hearing, each focusing on one of two speakers in a controlled setting. The auditory stimuli were dialogues between a male and a female speaker, Both native speakers, presented in various acoustic environments. Uniform loudness across the audio streams was achieved through RMS normalization, with the speakers positioned at 0° and 60° azimuths. Each session was designed to last 60 minutes, divided into 512 segments, with both speaker selection and stimuli sequence randomized in each trial. EEG data capture was performed using a 64-channel BioSemi ActiveTwo system, operating at a sampling rate of 512 Hz. Additional details regarding the DTU dataset are available in the cited studies~\citep{fuglsang2018eeg, fuglsang2017noise}.

\subsection{Performance of TAE Method}
To validate the superiority of our TAE method over alternative threshold-based encoding methods for raw signals, we conducted further tests focusing on three key metrics: the average mean absolute error (MAE) in signal reconstruction, the average spike firing rate, and the performance of our TAE method. As shown in Fig.\ref{picture6},
by counting the average Reconstruct-MAE of the three methods on the GSC and radar~\citep{} datasets, the TAE method was significantly smaller than the other two methods. The experimental outcomes reveal that the TAE method more closely approximates the original signals and more reliably preserves their critical features. 
\begin{figure}[htpb] 
    \centering
    \subfigure[]{
        \label{Fig.sub.qqq}
 	\includegraphics[scale=0.29]{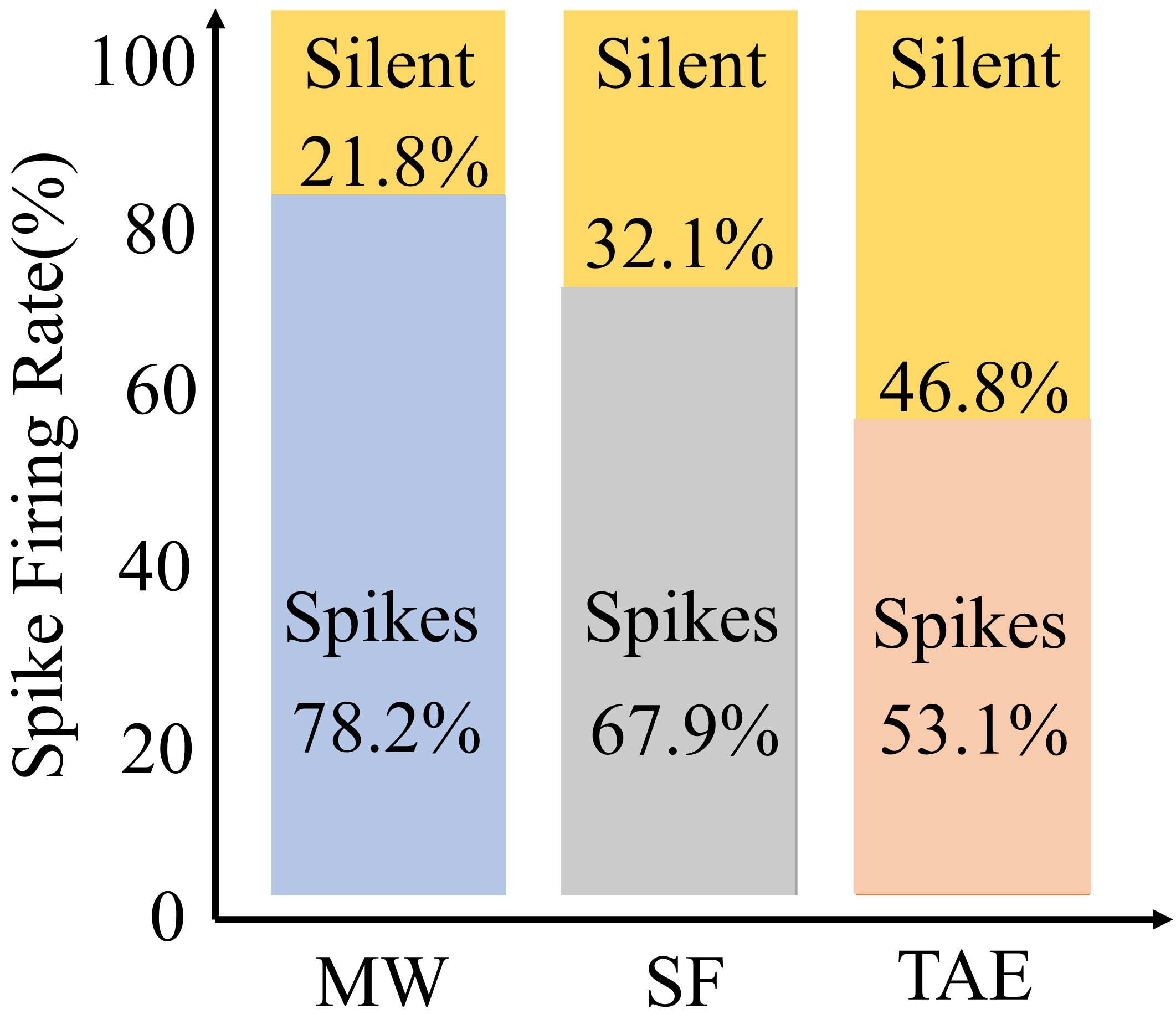}
   }  
   \subfigure[]{
        \label{Fig.sub.qqq}
 	\includegraphics[scale=0.29]{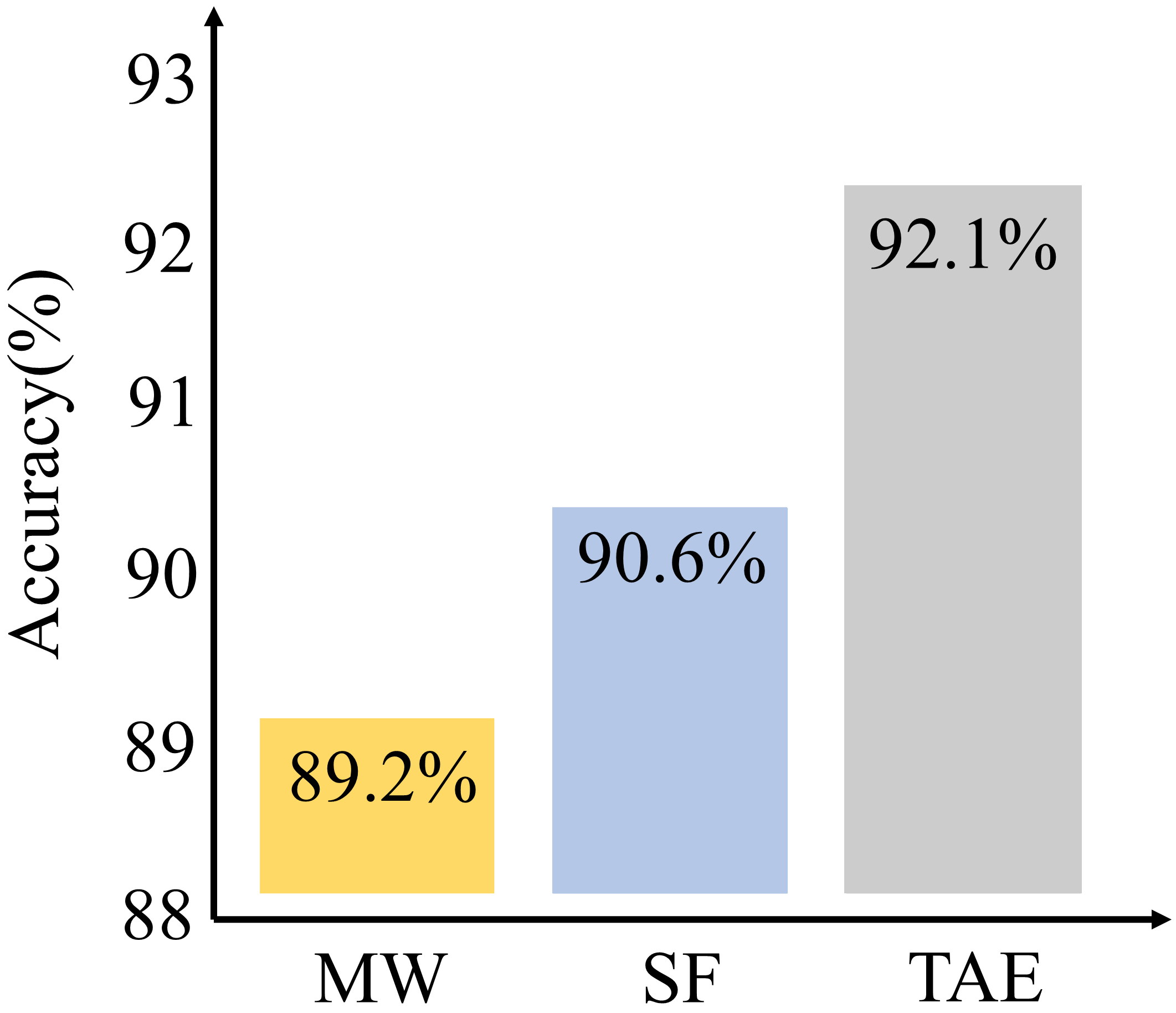}
   }  
    \caption{comparing TAE method against other encoding methods on performance and spike firing frequency. (a) Comparing respective accuracies on the GSC dataset among TAE, MW,  and SF under a uniform backend classification model. (b) The spike firing rates of TAE, SF, and MW on the GSC dataset.}
    \label{picture7}
\end{figure}
Additionally, as depicted in Fig.\ref{picture7}, 
by conducting tests using the same network architecture, we test the average spike firing rates and performance across these three encoding schemes on the GSC dataset.
The results demonstrate a lower peak firing rate and higher performance for the TAE method, indicating its ability to represent analog signals with better representation and more sparse spikes. Consequently, our TAE method can encode analog signals into ternary spikes more effectively at lower firing rate, significantly reducing the bandwidth requirements for signal transmission.

\subsection{Performance of QT-SNN}
\begin{figure*}[htpb] 
\centering
\includegraphics[scale=0.47]{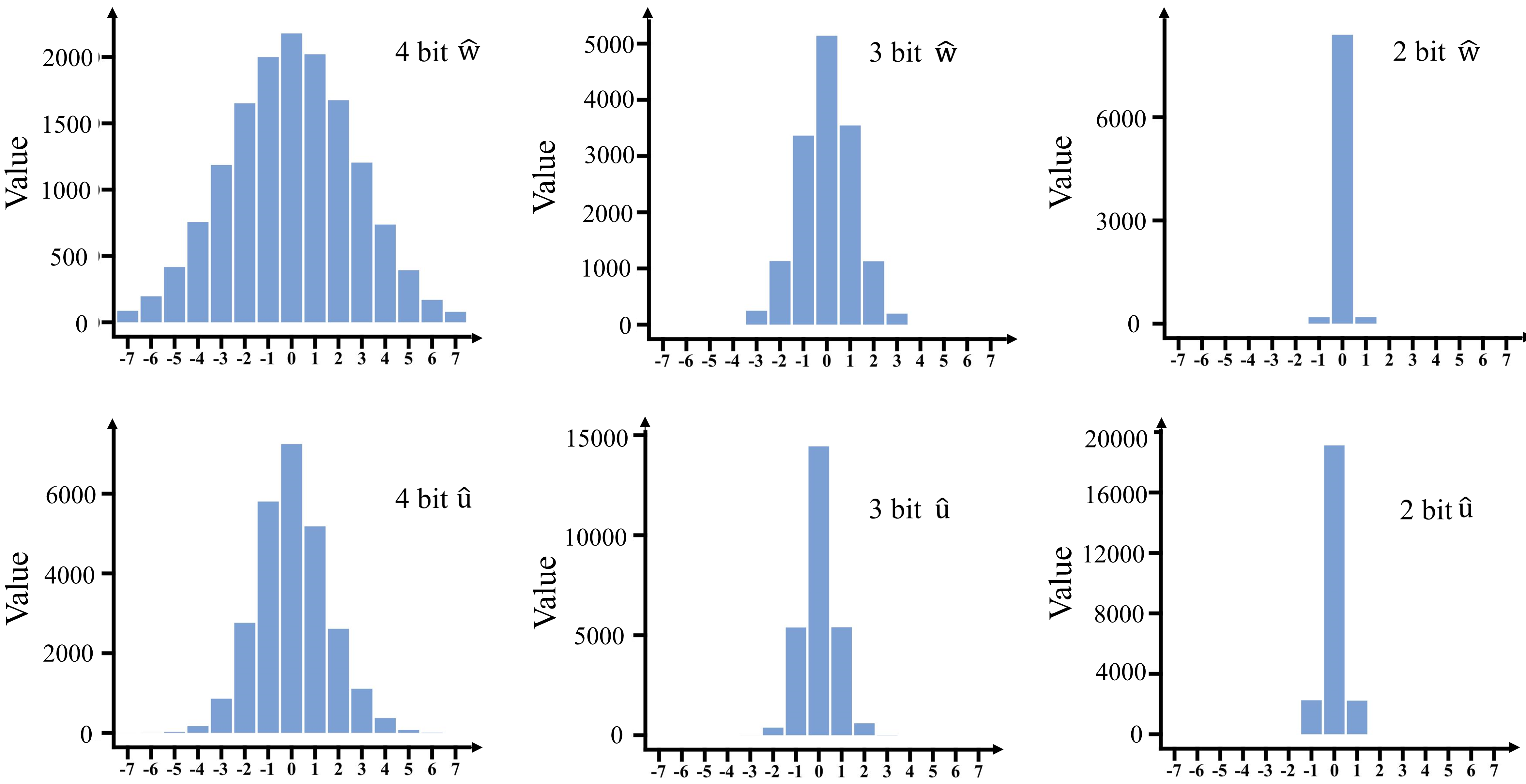}   

\caption{In our QT-SNN, when weights and membrane potentials are quantized to 4, 3, and 2 bits, their distributions adhere to normal distributions with varying standard deviations. Satisfactorily, as the bit-width decreases, the proportion of zero-valued weights (silent) and membrane potentials significantly increases, further enhancing the energy efficiency of the SNN.}
\label{picturebit}
\end{figure*}

To assess the performance of our QT-SNN, we compare it against other quantization methods on CIFAR10 with different network structures. As demonstrated in Table.\ref{lab1}, 
In the context of full precision for synaptic weights and membrane potentials, our QT-SNN demonstrates superior performance compared to other methods(MINT(\cite{yin2023mint}), TC-SNN(\cite{zhou2021ste}), CBP-QSNN(\cite{yoo2023cbp})). This is attributed to the fact that ternary spike neurons, alongside learnable parameters \(a_1/a_3\), can improve the information capacity, thereby enhancing the performance of QT-SNN.
Moreover, our QT-SNN exhibits no significant performance degradation at reduced bit widths for $w$ and $u$, underscoring the superiority of our quantization approach. Additionally, as illustrated in Fig.\ref{picturebit}, we analyzed the distribution of membrane potentials and weights across all layers at low bit-width levels, aligning with normal distributions of varying standard deviations. Notably, the proportion of zero values in membrane potentials and weights increases as the bit width decreases, further reducing the computational energy consumption of our QT-SNN.

\begin{table}[h]
\centering

  \caption{\centering{Comparison of QT-SNN with other methods}}
\begin{tabular}{lccccc}
\hline
Method & fp32 & \multicolumn{4}{c}{Precision (W/U)} \\
 & & 8/8 & 4/4 & 2/2 &1/32 \\
\midrule
ResNet-19 & & & & & \\
MINT & 91.29 & 91.36 & 91.45 & 90.79 &- \\
(Ours)QT-SNN & 94.59 & 94.31 & 93.99 & 93.71 &- \\
\midrule

VGG-16 & & & & & \\
TC-SNN &92.68 &- &- &- &91.51 \\
MINT & 91.15 & 90.72 & 90.65 & 90.56 &-\\
CBP-QSNN & 91.79 &- &- &- &90.93 \\
(Ours)QT-SNN & 94.27 & 93.88 & 93.56 & 93.28 &-\\

\midrule
VGG-9 & & & & \\

MINT  & 88.03 & 87.48 & 87.37 & 87.47 &- \\
(Ours)QT-SNN & 92.11 & 91.58 & 91.12 & 90.71 &-\\
\midrule
7Conv+2FC & & & & \\
CBP-QSNN  & 89.73 &- &- &- &89.01 \\
(Ours)QT-SNN & 91.27 & 90.88 & 90.90 & 90.75 &-\\
\midrule
\label{lab1}
\end{tabular}
\end{table}

\begin{table*}[t]
  \caption{\centering{Comparison of model performance on GSC datasets for 12 and 35 classification.}}
  \label{tab:kws}
  \centering
  \begin{tabular}{c c c c c c}
    \toprule
    \textbf{Model}  &  \textbf{Network}    &\textbf{Encoding}   &\textbf{Memory(MB)}   &\textbf{Precision(W/U)} &  \textbf{Acc(\%)}\\
    \midrule
    \multicolumn{6}{c}{Google Speech Commands Dataset Version 2 (12)}\\[1ex]
    ST-Attention-SNN \cite{wang2023spatial}     &SNN    &DC   &2170  &32/32 &95.1\\
    SLAYER-RF-CNN \cite{orchard2021efficient} & SNN    &DC   &888     &32/32 & 91.4  \\
    SpikGRU\cite{dampfhoffer2023leveraging}    & SNN     &DC  &1587    &32/32 & 92.9 \\

    (Our) SNN-KWS      &SNN   &DC    &863      &16/16 &94.5 \\
    (Our) SNN-KWS      &SNN   &TAE    &72      &16/16    &93.6 \\
    (Our) SNN-KWS      &SNN   &DC    &538      &4/4 &93.9 \\
    (Our) SNN-KWS      &SNN   &TAE    &63      &4/4 &93.2 \\
    (Our) SNN-KWS      &SNN   &DC    &527      &2/2 &93.4 \\
    (Our) SNN-KWS      &SNN   &TAE    &59      &2/2 & 92.9 \\
    \midrule
    \multicolumn{6}{c}{Google Speech Commands Dataset Version 2 (35)}\\[1ex]
    WaveSence \cite{weidel2021wavesense} &SNN      &DC  &N/A  &32/32 &79.5\\
    LSTMs-SNN \cite{zhang2023long}     &SNN      &DC  &N/A  &32/32 &91.5\\
    SRNN+ALIF \cite{yin2021accurate} & SNN     &DC  &3290     &32/32  & 92.5  \\
    Speech2Spikes\cite{stewart2023speech2spikes}   & SNN    &DC  &2780     &32/32  & 89.5 \\

    (Our) SNN-KWS      &SNN   &DC    &871    &16/16  &92.8 \\
    (Our) SNN-KWS      &SNN   &TAE    &74      &16/16 &92.3 \\
    (Our) SNN-KWS      &SNN   &DC    &853      &4/4 &92.3 \\
    (Our) SNN-KWS      &SNN   &TAE    &68      &4/4 &91.9 \\
    (Our) SNN-KWS      &SNN   &DC    &849      &2/2 &92.1 \\
    (Our) SNN-KWS      &SNN   &TAE    &60     &2/2 &91.8 \\
    \bottomrule
  \end{tabular}
\end{table*}

 \begin{table*}[t]
  \caption{\centering{Comparison of model performance on KUL and DTU datasets.}}
  \label{tab:eeg}
  \centering
\begin{tabular}{ccccccccccc}
\hline
\multirow{2}{*}{Dataset} & \multirow{2}{*}{Model} & \multirow{2}{*}{Network} & \multirow{2}{*}{Precision} & \multicolumn{7}{c}{Decision Windows(second)}    \\ \cline{5-11} 
                         &                       &                          &                            & 0.1  & 0.2  & 0.5  & 1    & 2     & 5    & 10   \\ 
    \midrule
\multirow{5}{*}{KUL}     & \cite{de2018decoding}           & ANN                      & 32                         & 50.9 & 53.6 & 55.7 & 60.2 & 63.5  & 67.4 & 75.9 \\
                         & \cite{vandecappelle2021eeg}        & ANN                      & 32                         & 74.3 & 78.2 & 80.6 & 84.1 & 85.7  & 86.9 & 87.9 \\
                         & \cite{cai2021eeg}                 & ANN                      & 32                         & 80.8 & 84.3 & 87.2 & 90.1 & 91.4  & 92.6 & 93.9 \\
                         & (Our) TAE+QT-SNN           & SNN                      & 4                         & 91.2 & 91.7 & 91.9 & 92.4 & 93.6 & 94.0 & 94.3 \\ 
\midrule
\multirow{5}{*}{DTU}     & \cite{de2018decoding}           & ANN                      & 32                         & -    & -    & -    & 53.4 & 57.7  & 61.9 & 70.1 \\
                         & \cite{vandecappelle2021eeg}       & ANN                      & 32                         & 56.7 & 58.4 & 61.7 & 63.6 & 65.2  & 67.4 & 67.8 \\
                         & \cite{cai2021eeg}                & ANN                      & 32                         & 65.7 & 68.1 & 70.8 & 71.9 & 73.7  & 76.1 & 75.8 \\
                    
                         & (Our) TAE+QT-SNN            & SNN                      & 4                         & 66.8 & 68.6 & 70.4 & 72.2& 73.6  & 76.8 & 76.6  \\
                         \hline

\end{tabular}
\end{table*}

\subsection{ Performance of Ternary Spike-based Neuromorphic Signal Processing System }
\begin{figure}[htpb] 
\centering
\includegraphics[scale=0.58]{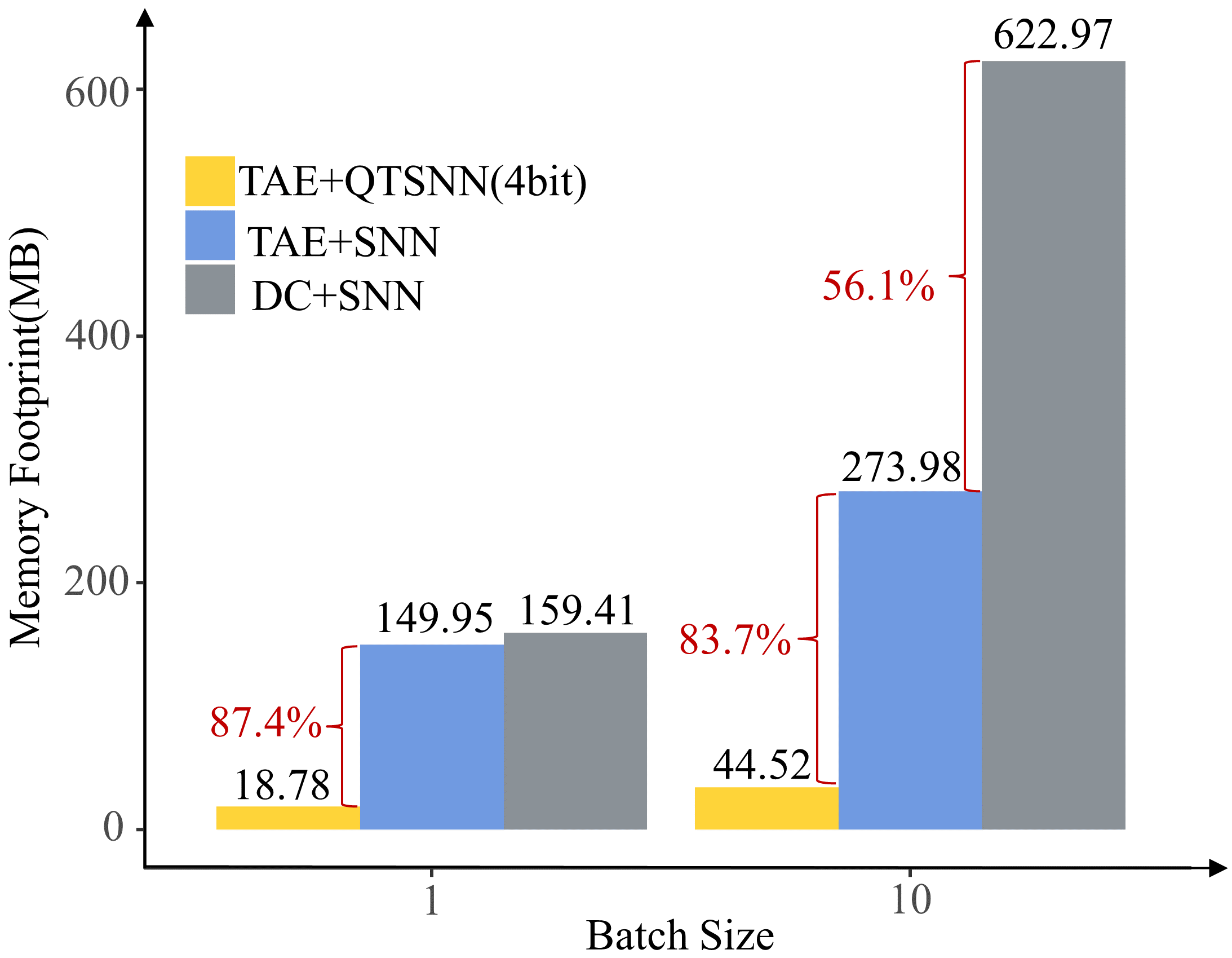}

\caption{Both TAE and QT-SNN contribute to memory reduction across varying batch sizes. With a batch size of 1, QT-SNN achieves approximately an 87.4\% reduction in memory usage. At a batch size of 10, TAE and QT-SNN reduce memory by 83.7\% and 56.1\%, respectively. Therefore, Both TAE and QT-SNN significantly contribute to reducing our system's memory footprints.}
\label{Fig.sub.aaaaaa}
\end{figure}

By integrating the strengths of the TAE method and QT-SNN, we design a more lightweight and energy-efficient ternary neuromorphic signal processing system.
To validate the performance and memory efficiency of our system, we establish numerous experiments across two classical signal processing tasks: keyword spotting and EEG recognition. As shown in Tables.\ref{tab:kws} and \ref{tab:eeg}, our system maintains SOTA performance with both membrane potentials and weights quantized to lower bit-width, outperforming other SNN-based signal processing solutions~\citep{weidel2021wavesense,zhang2023long,yin2021accurate,stewart2023speech2spikes,wang2023spatial,orchard2021efficient}. 
Regarding memory efficiency, we established ablation experiments to confirm that both the TAE method and QT-SNN contribute to reducing the system's memory footprint. As shown in Fig.\ref{Fig.sub.aaaaaa},
In scenarios with a batch size of 1, QT-SNN can achieve approximately 90\% reduction in memory usage, with negligible differences observed between DC and our TAE methods. However, for larger batch sizes, TAE encoding achieved up to 87\% memory reduction compared to the DC method. 
Consequently, although our TAE method exhibits a marginal performance decrement relative to the DC method, it substantially reduces the system's memory requirements, rendering it more suitable for deployment on resource-limited edge devices.
Moreover, we will detail the computational energy consumption advantages of our system through a theoretical energy consumption analysis in the subsequent section.

\subsection{Energy Efficiency}
To further validate the energy efficiency of our neuromorphic signal processing system, we calculated the theoretical energy consumption of Synaptic Operations (SynOps) for TAE+QT-SNN, DC+SNN, and the same structure ANN networks. According to the standards established in the field of neuromorphic computing(~\cite{sengupta2019going,liu2023low,liu2023aa}), the total SynOps required for the ANNs is defined as follows:
\begin{equation}
    EN_{(ANNs)} = MAC \times \sum_{l=1}^{L} f_{in}^l N_l,
\end{equation}
where \( f_{in}^l \) is the number of fan-in connections for the neurons in layer \( l \), \( N_l \) is the number of neurons in layer \( l \), and \( L \) is the total number of layers in the network.

The DC method, Widely used in SNNs, employs the original signal as the initial layer's input, repeating it across temporal steps.
Thus, the total synaptic operation energy consumption for the DC+SNN is defined as:
\begin{multline}
    EN_{(DC+SNN)} = \\
    MAC \times \sum_{t=1}^{T} \sum_{j=1}^{N_1} f_{in,j}^1 
    + AC \times \sum_{t=1}^{T} \sum_{l=2}^{L} \sum_{j=1}^{N_l} f_{out,j}^l o_j^l[t],
\end{multline}
where the left half of the formula represents the synaptic energy consumption of the DC encoding layer, and the right half represents the synaptic operations energy consumption of the subsequent spiking layers. Here, \( f_{out,j}^l \) is the number of fan-out connections from neuron \( j \) in layer \( l \), \( T \) is the analog time window, and \( o_j^l[t] \) is the spike occurrence of neuron \( j \) at time \( t \). Regrettably, the extensive use of MAC operations in the first layer significantly limits the energy efficiency of SNNs.

Our system achieves all ternary spike-based information transmission and computation, effectively avoiding MAC operations. Its’ total synaptic operation energy consumption is defined as:
\begin{equation}
    En_{(TAE+QT-SNN)}= AC \times \sum_{t=1}^{T} \sum_{l=1}^{L} \sum_{j=1}^{N_l} f_{out,j}^l o_j^l[t].
\end{equation}

Here, we evaluate the hardware energy costs of our system(TAE+QT-SNN), DC+SNN, and ANN on a speech recognition task using the same network architecture. Our network architecture consists of 4 D-conv1d layers, 2 bottleneck blocks, and 1 FC layer.
For SNN-based models, we conduct inference using $4$ time steps. The average spike sparsity for TAE+QT-SNN is $10.42\%$, while for DC+SNN, it is $18.27\%$ (excluding the first layer). ~\citep{guo2023ternary, hu2021spiking,horowitz20141, rueckauer2017conversion} indicates a MAC operation requires $4.6$pJ and an AC operation requires $0.9$pJ, we compile the hardware energy costs of the three models as demonstrated in Table.\ref{table:your_label}. The result reveals that our neuromorphic signal processing system offers an approximate $7.5\times$ and $12.25\times$ energy saving compared to similar DC-based
SNN model and ANN-based model with the same model structures. Our system significantly enhances the energy efficiency and hardware-friendliness of intelligent signal processing models.

\begin{table}[ht]
\centering
  \caption{\centering{Comparison of hardware energy consumption for different models}}
\begin{tabular}{ccccc}
\hline
Method      & MAC & AC & Timestep & Energy \\
\hline
Our & 0M  & 15.19M & 4 & 5.70uJ \\
DC+SNN & 1.86M & 13.33M & 4  & 42.99uJ \\
ANNs & 15.19 &0M & 1  & 69.87uJ \\
\hline
\end{tabular}
\label{table:your_label}
\end{table}

\section{Conclusion}
This study introduces a ternary spike-based neuromorphic signal processing system to achieve lightweight, energy-efficient, and high-performance intelligent signal processing models for resource-constrained edge devices. This approach effectively addresses the challenges of neural signal encoding and model complexity inherent in existing SNN-based solutions. Our system incorporates two innovative components: Firstly, we introduce the TAE method, which more efficiently encodes raw analog signals into ternary spike trains, facilitating more sparse information transmission. Secondly, our QT-SNN model complements our TAE method, achieving direct processing of ternary spike signal information. Moreover, it quantizes both synaptic weights and membrane potentials to lower bit-width, significantly reducing the memory footprint and enhancing the hardware friendliness of intelligent signal processing models. Extensive experimental evidence demonstrates that our system, in comparison to other similar SNN-based works, achieves superior performance with greater lightweight and energy-saving characteristics. Future work will focus on deploying our system on neuromorphic chips, promising to offer a novel perspective for edge signal processing.

\printcredits

\bibliographystyle{model5-names}

\bibliography{my}

\end{document}